\documentclass[twocolumn,nofootinbib]{revtex4-2}
\usepackage[modulo]{lineno}

\usepackage{graphicx}  

\usepackage[T1]{fontenc}
\usepackage{dcolumn}   

\usepackage{bm}        
\usepackage{amsfonts}  
\usepackage{amsmath}   
\usepackage{amssymb}   
\usepackage{upgreek}

\usepackage{hyperref}

\usepackage[capitalise]{cleveref}

\newcommand{\dd}{\mathrm{d}}
\newcommand{\e}{\mathrm{e}}

\newcommand{\pwisein}{\left\{ \begin{array}{ll}}
\newcommand{\pwiseout}{\end{array}\right.}

\def\eV{\text{eV}}

\def\Xmax{X_\text{max}}

\def\gcm{\text{g}\,\text{cm}^{-2}}

\def\eV{\text{eV}}
\def\MeV{\text{MeV}}

\begin{document}

\title{The Greisen Function and its Ability to Describe Air-Shower Profiles}

\author{Maximilian Stadelmaier}
\affiliation{\it Institute of Physics of the Czech Academy of Sciences, Prague}
\author{Vladim\'ir Novotn\'y}
\affiliation{\it Institute of Physics of the Czech Academy of Sciences, Prague}
\affiliation{\it Institute of Particle and Nuclear Physics, Faculty of Mathematics and Physics, Charles University, Prague}
\author{Jakub V\'icha}
\affiliation{\it Institute of Physics of the Czech Academy of Sciences, Prague}

\date{March 2023}

\begin{abstract}  

\noindent
Ultrahigh-energy cosmic rays are almost exclusively detected through extensive air showers, which they initiate upon interaction with the atmosphere.
The longitudinal development of these air showers can be directly observed using fluorescence detector telescopes, such as those employed at the Pierre Auger Observatory or the Telescope Array.
In this article, we discuss the properties of the Greisen function, which was initially derived as an approximate solution to the electromagnetic cascade equations, and its ability to describe the longitudinal shower profiles.
We demonstrate that the Greisen function can be used to describe longitudinal air-shower profiles, even for hadronic air showers.
Furthermore we discuss the possibility to discriminate between hadrons and photons from the shape of air-shower profiles using the Greisen function.
\end{abstract}

\maketitle

\section{Introduction}

Extensive air showers are created by cosmic rays upon interaction with the atmosphere \citep{rossi1933eigenschaften,Auger:1939sh}.
They can be detected at the ground using surface detector arrays, or directly observed at night using fluorescence detector telescopes. 
To reconstruct the shower development and shower observables, a profile function needs to be fitted to the detector data.
Gaisser and Hillas proposed an empiric function \cite{gaisser1977reliability} to describe the longitudinal development of proton air showers as an alternative to the Constant Intensity Cut method \citep{Hersil:PhysRevLett.6.22,Alvarez-Muniz:2002tmp}, which is used in surface detector experiments to take into account the atmospheric attenuation of particles in air showers from different zenith angles.
It was shown in \cite{Montanus:2011hg} that the Gaisser--Hillas (GH) function can be used to approximate a system of particles being created and absorbed in an extended Heitler--Matthews model \citep{Matthews:2005sd}, and can be adjusted to very closely match the Greisen function, which is an approximation for the solutions to the electromagnetic cascade equations \citep{Greisen:1960}.
Both the Pierre Auger Observatory and the Telescope Array use the GH function to describe their fluorescence detector data \citep{PierreAuger:2014sui,TelescopeArray:2018xyi}.

In this article, we will discuss the Greisen function and its properties, such as its connection to the shower age.
We will demonstrate the usability of the Greisen function as an alternative to the GH function to fit longitudinal shower profiles and present its performance to reconstruct the depth of the shower maximum as well as the primary energy using Monte Carlo (MC) simulations of air showers.

\section{The Greisen Function}

The average longitudinal development of electromagnetic air showers can be very well described analytically \citep{Rossi:1941zza}.
This description holds in good approximation also for hadronic showers, initiated by ionized nuclei, which make up the upper end of the cosmic-ray energy spectrum \citep{Kampert:2012mx}.
The solutions to the cascade equations derived by Rossi and Greisen under Approximation A\footnote{Approximation A is the high-energy approximation to the cascade equations, in which only bremsstrahlung and pair production are considered as relevant processes.} to describe extensive air showers were used to motivate important properties in the context of air-shower physics, such as the shower age
\begin{align}
    s = \frac{3t}{t + 2\ln(E_0/E_\text{cut})}
    \label{eq:s}
\end{align}
that describes the development of an electromagnetic shower, initiated by a primary particle of the energy $E_0$, after $t$ radiation lengths, considering only the particles above an energy of $E_\text{cut}$.
Note that $s=1$ at $t = \ln(E_0/E_\text{cut})$; this value is usually assigned with the shower maximum.
For electromagnetic showers a reasonable choice for $E_\text{cut}$ is close to ${\approx}87\,\MeV$, which is the energy above which electromagnetic particles on average lose more energy in radiative shower processes than to scattering and ionization.
Furthermore, it was demonstrated that the relative rate of change\footnote{We use the notation $\lambda_1$, even though we do not mean to suggest that this quantity is to be understood as a (wave) length, to adhere to historic convention.} $\lambda_1$ in the number $N$ of particles as a function of the surpassed radiation lengths $t$, 
\begin{align}
    \lambda_1 = \frac{1}{N(t)}\frac{\partial N(t)}{\partial t},
    \label{eq:lambda_1}
\end{align}
is similar for all electromagnetic showers at high primary energies.
Greisen introduced the approximation \cite{Greisen:1960}
\begin{align}
    \lambda_1 \simeq \frac{1}{2}\left(s-1-3\ln s\right),
    \label{eq:lambda_1_approx}
\end{align}
which is in good agreement with the exact solution.
Originally, the parameter $s$ describes the spectra $n$ of electromagnetic particles in a shower, that is approximately given by $n_\upgamma \sim n_{\text{e}^\pm}\sim E^{-(s+1)}$ for particles at energies $E\ll E_0$, but from \cref{eq:lambda_1_approx} there exists a relation between $\lambda_1$ and $s$.

It is straightforward to combine \cref{eq:s,eq:lambda_1,eq:lambda_1_approx} and to solve the resulting expression for $N(t)$ by integration.
This yields
\begin{align}
    N(t) = N_0 \exp{\left[t\left(1-\tfrac{3}{2}\ln s\right)\right]},
    \label{eq:greisen_unfinished}
\end{align}
with a constant $N_0$.
The maximum number $N_\text{max}$ of particles above the energy of $E_\text{cut} = 98\,\MeV$ in a cascade initiated by a particle of energy $E_0$ was derived in \cite{Belenky:1946} under Approximation B\footnote{Approximation B of the cascade equations augments Approximation A by a term concerning Coulomb scattering.} and found to be
\begin{align}
    N_\text{max} = \frac{0.31}{\sqrt{\ln(E_0/E_\text{cut})}} \frac{E_0}{E_\text{cut}}.
    \label{eq:tamm_belenky}
\end{align}
\cref{eq:greisen_unfinished} has its maximum at $t_\text{max}=\ln(E_0/E_\text{cut})$, which evaluates to $N_0\, E_0/E_\text{cut}$.
Thus, solving for $N_0$ using \cref{eq:tamm_belenky} yields the Greisen function, which reads as
\begin{align}
     N(t) = \frac{0.31}{\sqrt{\beta}}\exp{\left[t\left(1-\tfrac{3}{2} \ln s\right)\right]},
     \label{eq:greisen_fct}
\end{align}
using the short notation $\beta = \ln(E_0/E_\text{cut}) = t_\text{max}$.
The Greisen function, introduced in \citep{1956progress}, is thus an approximate solution to the electromagnetic cascade equations, given in \citep{Rossi:1941zza}, combining aspects of both Approximation A and Approximation B.
There was no strict derivation given by Kenneth Greisen himself, but \emph{a-posteriori} derivations (such as the one presented here) were provided in \cite{Lipari:2008td} and \cite{Stadelmaier:2022tbt}.

In the following, we have to overcome two major shortcomings of the Greisen function as written in \cref{eq:greisen_fct}.
Firstly, the Greisen function in its classical form cannot accurately describe the point of the first interaction of a cosmic ray with the atmosphere, since by construction the cascade is initiated always at $t=0$.
Secondly, the scale of the Greisen function is only accurate for average electromagnetic showers.
We will introduce a parameter $\epsilon$ to account for this issue and demonstrate that the Greisen function generalized this way is able to describe the longitudinal profile of hadronic showers and the corresponding shower-to-shower fluctuations.

\section{The Modified Greisen Function}

The classical Greisen function, which is given in \cref{eq:greisen_fct}, assumes that a shower starts at $t=0$.
Furthermore, the Greisen function is technically only able to describe electromagnetic showers. 
In this section, we introduce minor modifications to the function to describe the longitudinal development of both hadronic and electromagnetic air showers.

Firstly, we introduce a non-zero point of the first interaction at a slanted atmospheric depth $X_1$, that will be described by $t_1 = X_1/X_0$ using the electromagnetic radiation length\footnote{Equivalently, we use $t=X/X_0$ and $t_\text{max}= \Xmax / X_0$.} $X_0 \simeq 37\,\gcm$.
Thus, the shower age $s$ will be given as
\begin{align}
    s = \frac{3t^\prime}{t^\prime + 2 \beta} \,\Theta (t^\prime),
    \label{eq:smod}
\end{align}
with $t^\prime=t-t_1$ and the Heaviside function $\Theta$.
To maintain the property of the shower age, which is supposed to be 1 at the maximum of the shower, and to keep number of radiation lengths required to reach the maximum of the shower the same as before, we redefine $\beta$ in accordance with the previous modification as
\begin{align}
    \beta = \ln(E_0/E_\text{cut}) = t_\text{max} - t_1.
    \label{eq:beta}
\end{align}
Finally, we introduce the factor $\epsilon$, which is defined in units of energy deposit per step length, and which can be interpreted as the effective energy loss per particle and step length at the shower maximum\footnote{Here we ignore the factor of 0.31 from \cref{eq:tamm_belenky}.}.
Thus, the modified Greisen profile reads as

\begin{align}
    N(t) \equiv \frac{\dd E}{\dd X}(t) = \frac{\epsilon}{\sqrt{\beta}} \exp\left[{(t-t_1)\left(1-\tfrac{3}{2}\ln s\right)}\right],
    \label{eq:mGreisen}
\end{align}
with $N(t)=0$ for $t\leq t_1$.
For the sake of simplicity, here and in the following, in the text we abbreviate the energy deposit $\dd E / \dd X$ with the symbol $N$, analogously to the number of particles.

\section{Calibration of the Greisen Profile}
\label{sec:calib}

If \cref{eq:mGreisen} is used to describe the longitudinal profiles of (hadronic) showers in terms of deposited energy rather than a number of particles, it is necessary to examine viable (effective) numerical values of the energy $E_\text{cut}$ as well as of $\epsilon$.

We investigate the behaviour of $\epsilon$ and $E_\text{cut}$ using the MC values of $\Xmax$, $X_1$, $E_0$, and the maximum energy deposit $N(t_\text{max})$ of the longitudinal profiles of simulated air showers.
The simulations were produced using the \textsc{Sibyll 2.3d} \citep{Riehn:2017mfm}, \textsc{Epos-LHC} \citep{Pierog:2013ria}, and the \textsc{Qgsjet II-04} \citep{Ostapchenko:2010vb} models of hadronic interactions with different primary particles at different primary energies.
All simulations were produced in the \textsc{Conex} event generator \citep{Bergmann:2006yz} at version v7.60.
We produced 1000 simulated showers at primary energies of $10^{18.5}\,\eV$, $10^{19}\,\eV$, and $10^{19.5}\,\eV$ with gamma-ray, proton, and iron nuclei primary particles, each.

Shower-to-shower fluctuations of (hadronic) air showers will severely affect the maximum number of particles produced in a shower as well as the absolute depth of the shower maximum.
These fluctuations, however, can be accurately reproduced by the behaviour of the Greisen function.
Rewriting \cref{eq:beta} in terms of the slanted atmospheric depth of the shower maximum $\Xmax$ and the average radiation length $X_0$,
\begin{align}
    \beta = \frac{\Xmax-X_1}{X_0},
    \label{eq:beta2}
\end{align}
we find that
\begin{align}
    E_\text{cut} = E_0\,\e^{-\beta}= E_0 \exp{\left[-\frac{\Xmax-X_1}{X_0}\right]}.
    \label{eq:ecut}
\end{align}
Note that the explicit dependence of $E_\text{cut}$ on $E_0$ is cancelled by the dependence of the average $\Xmax$ on the logarithm of the primary energy in \cref{eq:beta}.
We choose an effective value for the radiation length of $X_0=40\,\gcm$ as a compromise for the different considered primary particles.

In a similar manner, using the numerical maximum of a given shower profile according to \cref{eq:mGreisen}, the parameter $\epsilon$ can be identified as 
\begin{align}
    \epsilon = N(t_\text{max})\,\sqrt{\beta}\,\e^{-\beta}.
    \label{eq:eps}
\end{align}

\begin{figure}
    \centering
    \vspace{1em}
    \includegraphics[width=0.9\columnwidth]{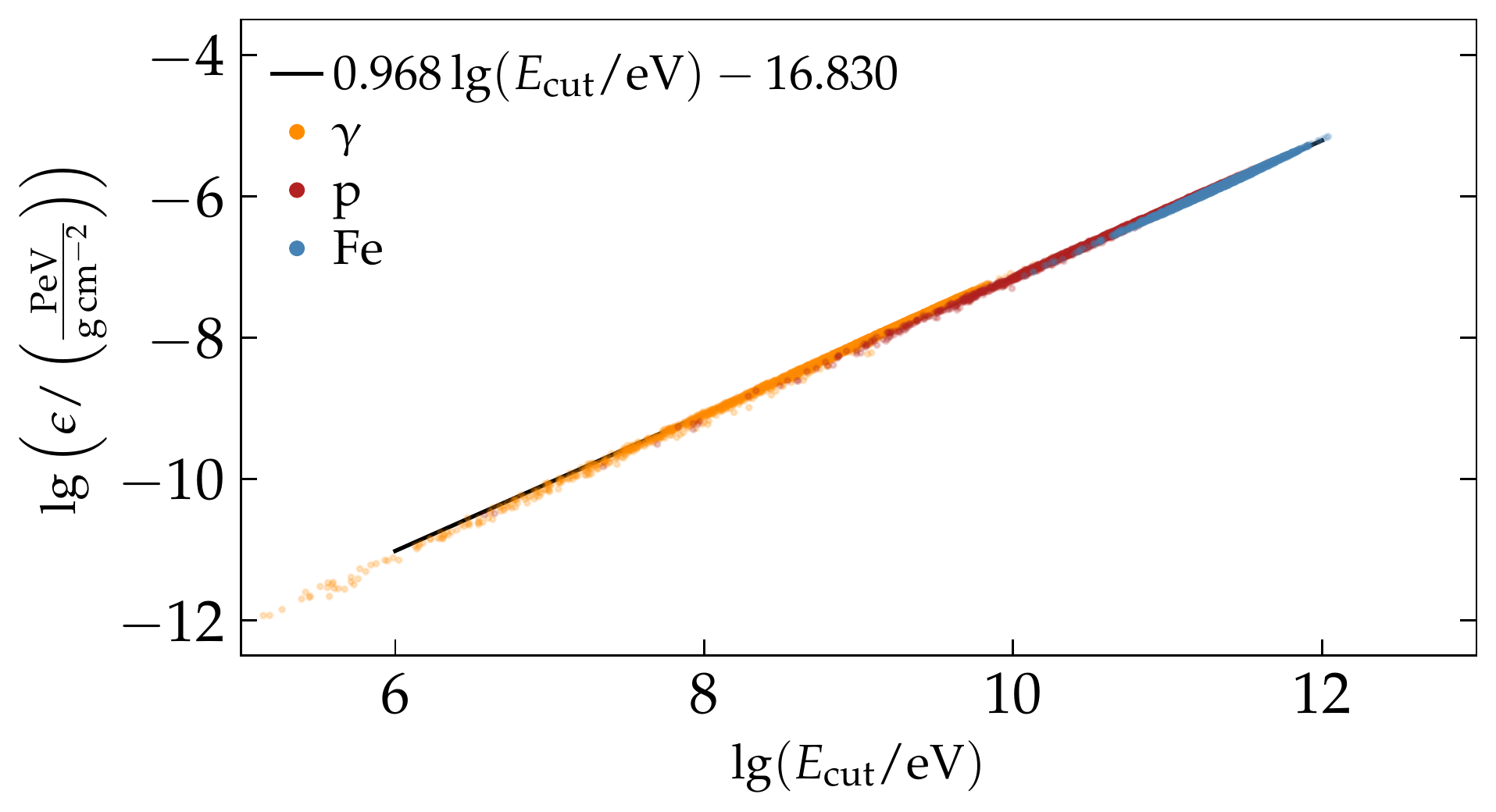}
    \caption{The behaviour of $\epsilon$ and $E_\text{cut}$ calculated from the Monte-Carlo values of $X_1$, $\Xmax$, $N(t_\text{max})$, and $E_0$ from simulated air showers produced with the \textsc{Sibyll2.3d} model of hadronic interactions.
    $E_\text{cut}$ and $\epsilon$ were obtained using \cref{eq:ecut} and \cref{eq:eps}, respectively. 
    The markers show data points from individual shower simulations, the line shows a log-linear regression.}
    \label{fig:ecut_eps_corr}
\end{figure}

As can be seen in \cref{fig:ecut_eps_corr}, we observe a strong correlation for the MC values of $\epsilon$ and $E_\text{cut}$ (and thus $\beta$), assuming Greisen-like longitudinal profiles.
Furthermore, we show in \cref{fig:calib_all_hims} and \cref{fig:calib_eps_ecut_values} that the distributions of $\epsilon$ and $E_\text{cut}$ as well as their relation are approximately the same even for all three hadronic interaction models.
This behaviour indicates that shower-to-shower fluctuations of hadronic showers are not at all random, but follow certain regularities. 
For example, a deeper than average value of $\Xmax$ corresponds to smaller than average value of $N(t_\text{max})$ (and vice-versa), if all other parameters are fixed.
Furthermore, we expect larger values of $\beta$ for photon-induced showers than for protons or iron nuclei.
Between different primaries there is a gradual transition in the shape of the shower from very hadronic (iron-like) to proton-like and lastly electromagnetic showers, where the behaviour of $\epsilon$ and $E_\text{cut}$ can be described by a power law, 
\begin{align}
    \frac{\epsilon}{\text{PeV}/\gcm} \simeq \left(\frac{E_\text{cut}}{10^{16.8}\,\eV}\right)^{0.97},
    \label{eq:eps_ecut_relation}
\end{align}

 with residuals on average within 0.2\%.
Note that if expressed in terms of $\Xmax$ and $X_1$ (cf.\ \cref{eq:ecut}), $E_\text{cut}$ is not an explicit parameter of the Greisen function.
\cref{eq:eps_ecut_relation} thus expresses the universal relation between the maximum energy deposit and the extent of the shower in terms of radiation lengths after the first interaction.

Even though $\epsilon$ appears as a pre-factor in the modified Greisen function, the numerical values of $\epsilon$ from a best fit are independent of the integrated profile (cf.\ \cref{eq:eps_ecut_relation,eq:ecut} and \cref{fig:calib_eps_ecut_values}) and thus of the energy of the primary particle.
Because $E_\text{cut}$ does not depend on the primary energy, $\epsilon$ solely depends on the shape of the shower.
And thus, as can be seen in \cref{fig:ecut_eps_corr}, depends on the amount of hadronization occuring during the shower development, which relates to the primary mass.
The integrated profile (and thus the calorimetric energy deposit) is mainly governed by the value of $\beta$ (cf.\ \cref{eq:integrated}).
For smaller values of $\epsilon$ (e.g.\ photon-like showers), the shower takes longer to reach its maximum in terms of radiation lengths.
In case of showers with a significant amount of hadronization, where multiple cascades are effectively in superposition (i.e.\ iron-like showers), we expect larger numerical values for $\epsilon$, corresponding to an earlier shower maximum.
From superposition and the Heitler--Matthews model\footnote{The model stimates the difference of the averages in $\Xmax$ for proton and iron showers to be $\langle{X}_\text{max}\rangle_\text{p} - \langle{X}_\text{max}\rangle_\text{Fe} \simeq 150\,\gcm$.} \citep{Matthews:2005sd} one would expect $\epsilon_\text{Fe} /\epsilon_\text{p} \simeq 50$ (cf.\ \cref{eq:mGreisen,eq:eps}); however, the average ratio obtained from simulations is much smaller.
From the ratio $\epsilon_\text{Fe} /\epsilon_\text{p} \simeq 10$, we estimate a difference in $\beta$ of about 2.4 radiation lengths between the average proton and iron shower to reach the shower maximum.
This is in accordance with simulations, which imply\footnote{The effect of the depth of the first interaction is neglected.} $\langle{X}_\text{max}\rangle_\text{p} - \langle{X}_\text{max}\rangle_\text{Fe} \simeq 100\,\gcm$.

\section{Fitting Simulated Data}
\label{sec:fit}

\begin{figure}
    \centering
    \vspace{1em}
    \includegraphics[width=0.98\columnwidth]{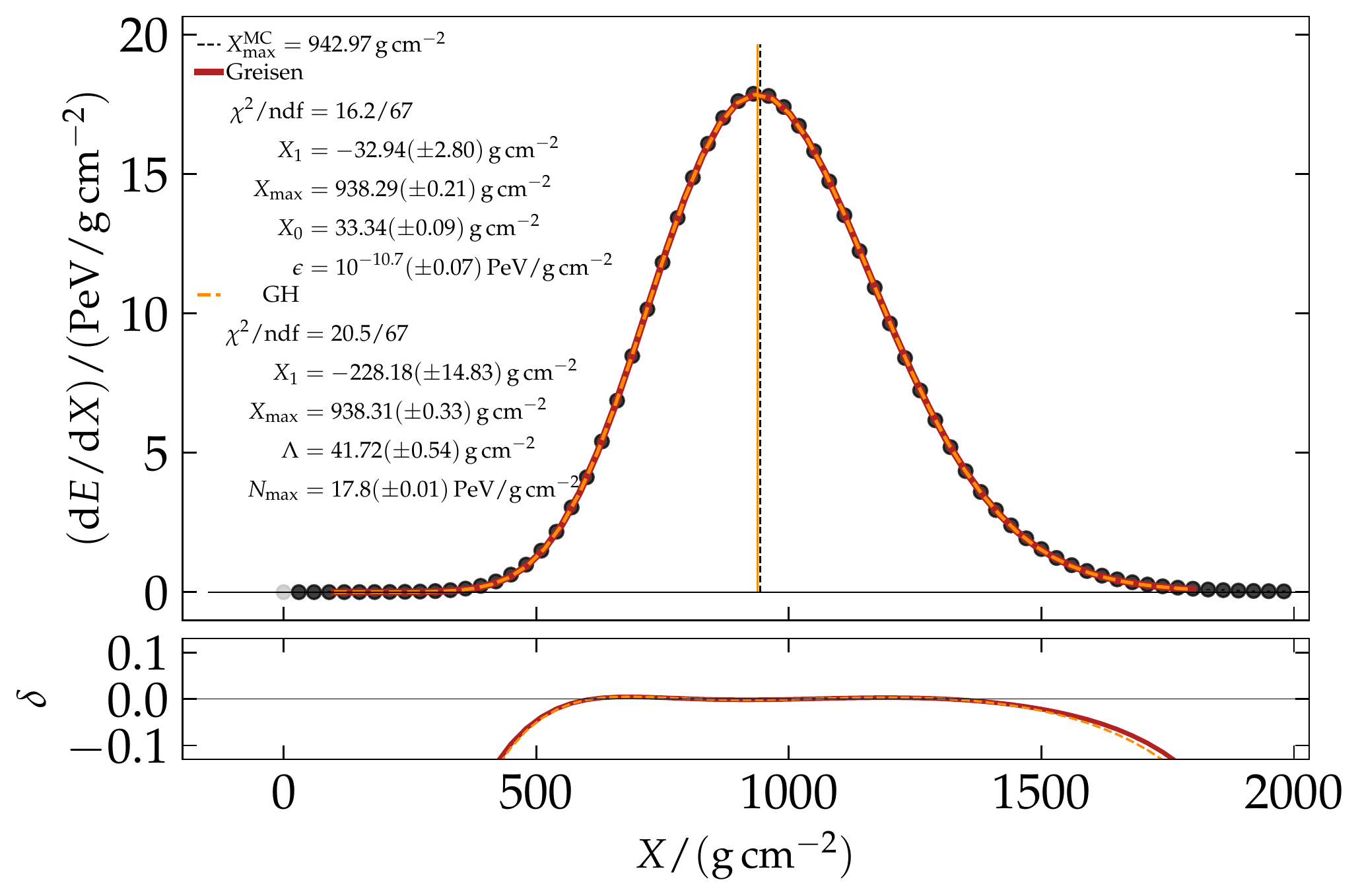}\\[0.5em]
    \includegraphics[width=0.98\columnwidth]{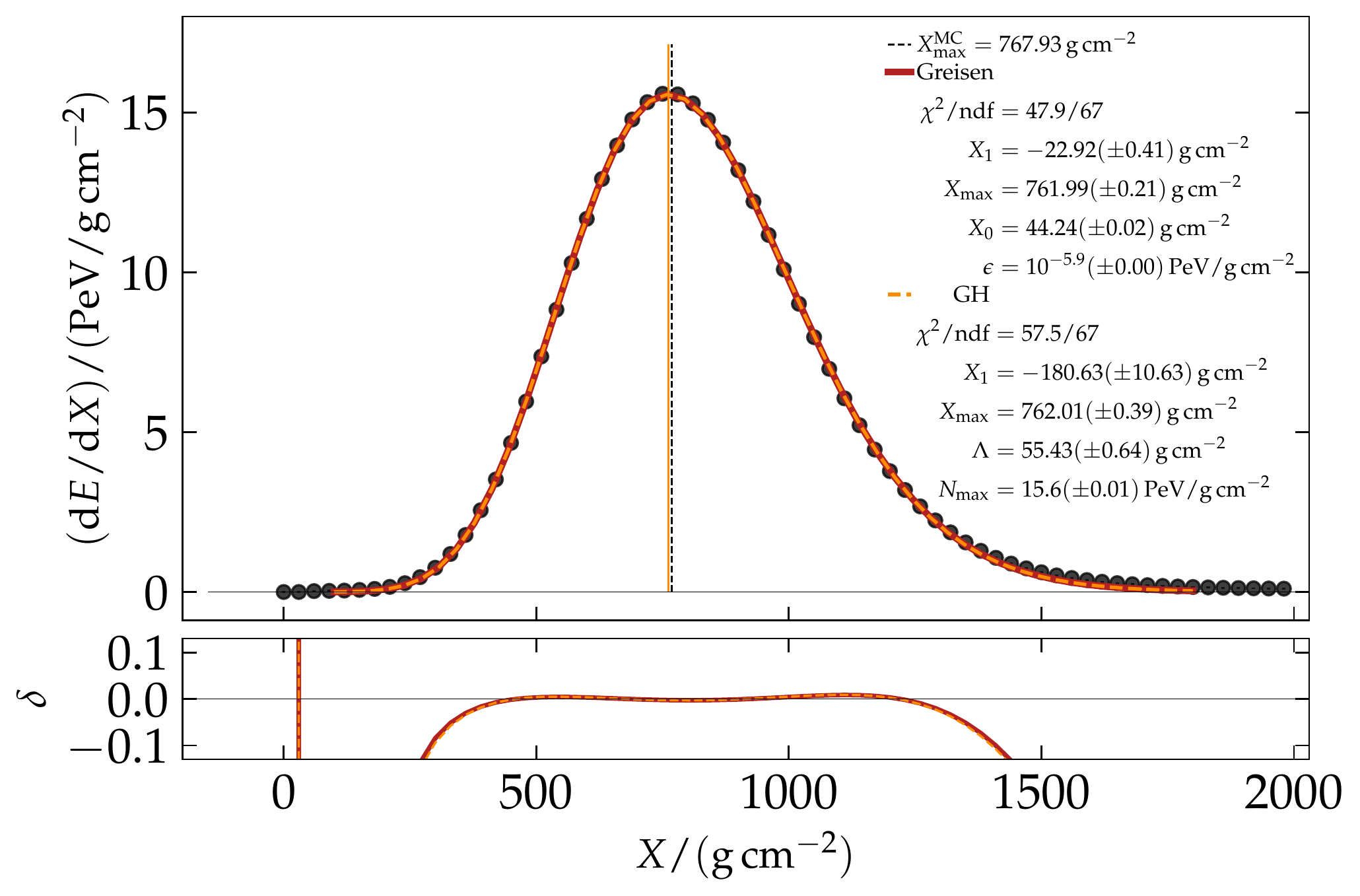}\\[0.5em]
    \includegraphics[width=0.98\columnwidth]{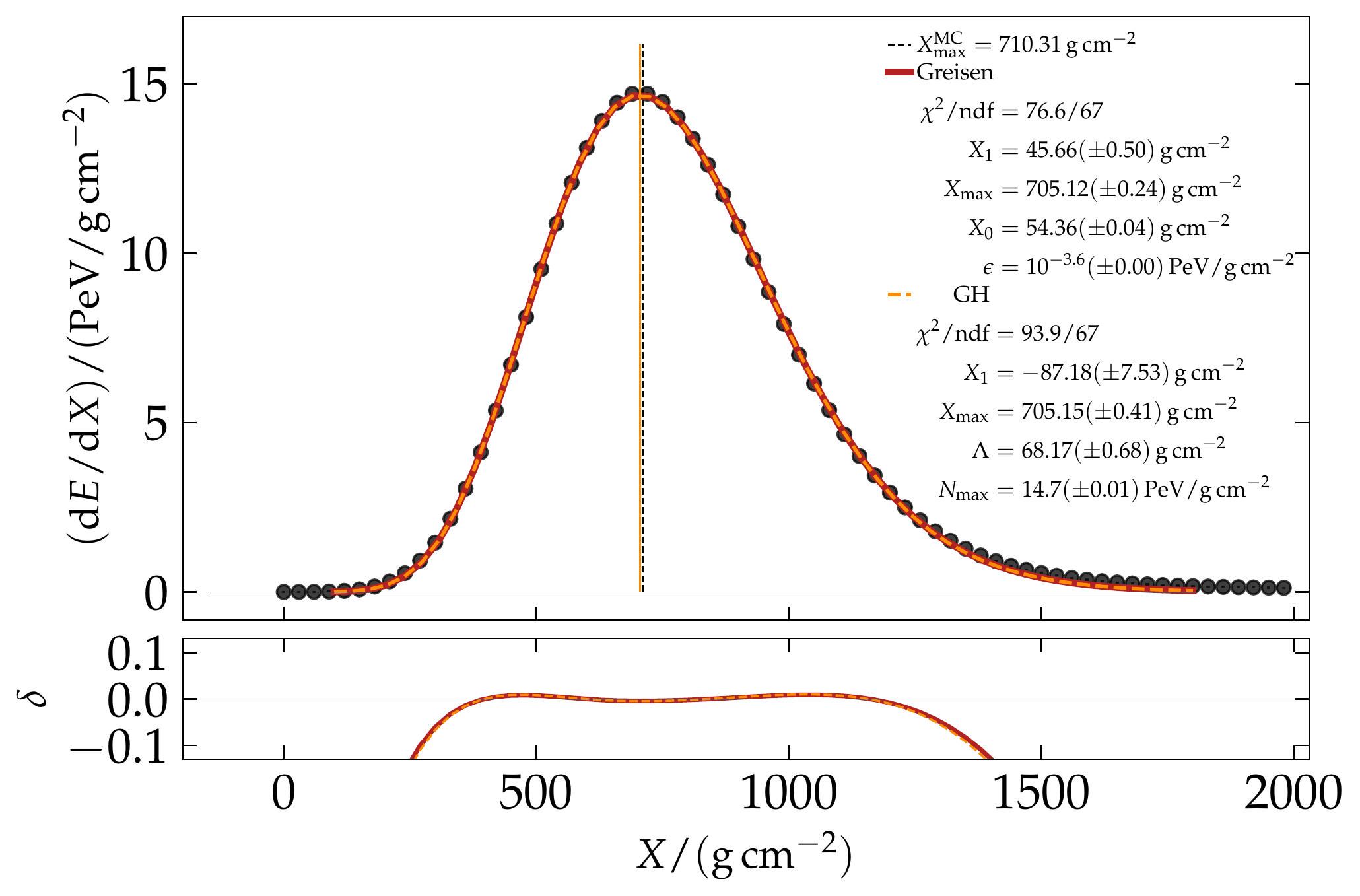}
    \caption{Example shower profiles with corresponding best fits.
    The showers were initiated by a (\emph{top to bottom}) gamma-ray, proton, and iron primary particle with a primary energy of $10^{19}\,\eV$ each, using the \textsc{Sibyll2.3d} model of hadronic interactions.
    The best fit values corresponding to both the Greisen function (\emph{red}) and the Gaisser-Hillas function (\emph{orange dashed}) are given in each panel along with the MC values of $\Xmax$. Additionally, the best fit values of $\Xmax$ are depicted by vertical lines. The profile tails (\emph{gray data points}) were disregarded for the fits.
    Below each panel, the relative deviation $\delta = (f - d)/d$ of the function value $f$ and the simulated profile data $d$ is shown for both functions in the respective color.}
    \label{fig:examples}
\end{figure}

\begin{figure}
    \centering
    \vspace{1em}
    \includegraphics[width=0.9\columnwidth]{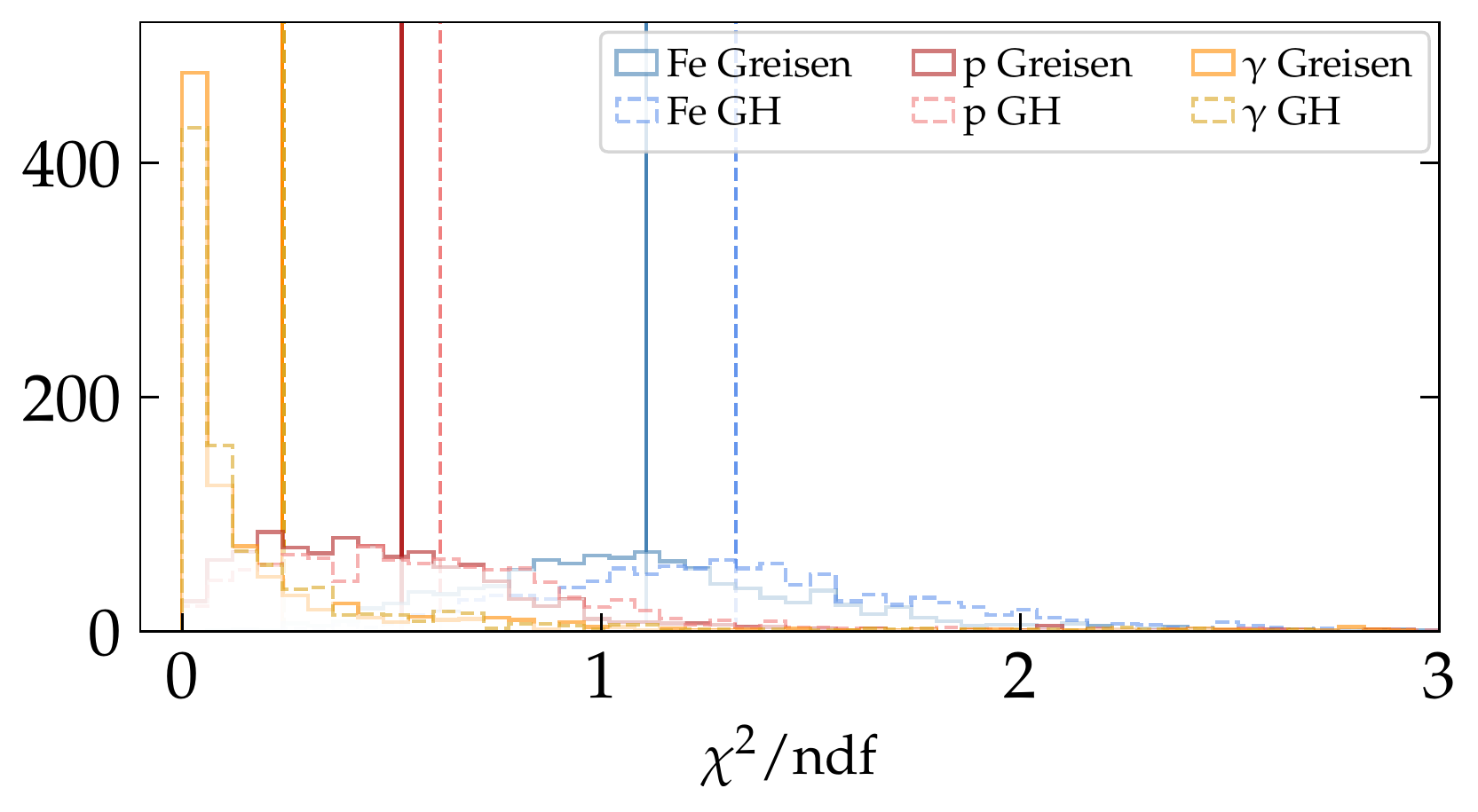}
    \caption{$\chi^2$-distributions for the Greisen and the GH functions fitted to simulated \textsc{Sibyll2.3d} air showers with primary energies of $10^{19}\,\eV$.
    The distributions for the Greisen (GH) functions are shown as a full (dashed) line, for each of the three primary particles.
    The respective mean of each distribution is indicated by a vertical line.
    }
    \label{fig:chisq}
\end{figure}

To test the ability of the Greisen function to describe longitudinal profiles of air showers, we examine and fit simulated shower profiles.
The simulation library contains the same configuration of primary energies, particles, and hadronic interaction models as mentioned before.
We compare the results against results from fitting the same showers to the very commonly used GH function, which in terms of the slanted depth $X$ reads as
\begin{align}
    \begin{split}
    N(X) = N_\text{max} & \left(\frac{X - X_1}{\Xmax - X_1} \right)^{\tfrac{\Xmax-X_1}{\Lambda}} \\
    & \hspace{5em} \times \exp{\left[\frac{\Xmax - X}{\Lambda}\right]},
    \end{split}
    \label{eq:gh}
\end{align}
with $N(X) = 0$ for $X \leq X_1$, the depth of the shower maximum $\Xmax$, the maximum value $N_\text{max}$ of the function, and a characteristic length $\Lambda$, which is related but not equal to the electromagnetic interaction length.
Approximately one expects $\Lambda \simeq 3\,X_0/2$ (cf.\ \cref{eq:greisen_alternative} and \cite{Lipari:2008td,Stadelmaier:2022tbt}).

To describe the shape of the shower profiles, we use $N_\text{max}$, $\Xmax$, $X_1$, and $\Lambda$ as free parameters for the GH function, and $\epsilon$, $\Xmax$, $X_1$, and $X_0$ for the Greisen function.
We estimate the uncertainty of the individual MC data points to $0.3\,\text{PeV}/(\gcm)$.
The tails of the profiles ($\dd E/\dd X \leq 0.8\,\text{PeV}/(\gcm)$) are not used in any of the fits.
The constant threshold for the tails as well as the uncertainty for the data points were estimated so that $\chi^2/\text{ndf}$ is  $\sim1$ for a mixture of proton and iron showers with primary energies of $10^{19}\,\eV$.
Examples of fitted profiles  of the photon, proton, and iron nucleus induced showers are given in \cref{fig:examples}.

The $\chi^2$-distributions for the Greisen and GH functions fitted to simulated data from \textsc{Sibyll2.3d} showers at a primary energy of $10^{19}\,\eV$ are depicted in \cref{fig:chisq}.
Additional $\chi^2$-distributions from showers simulated with different hadronic interaction models and primary energies are given in \cref{fig:chisq_additional} (note that the estimated uncertainty of the individual data points, as well as the limit for fitting the tails was not adjusted for the different primary energies).
We find that on average for all energies and hadronic interaction models, the average values of $\chi^2/\text{ndf}$ are smaller for the Greisen function fit, than for the fit using a GH function, thus implying a better match of the function to the profile data.
For iron showers the difference of $\chi^2/\text{ndf}$ is the largest, with values being approximately $20\%$ smaller for the Greisen function.
The difference for proton showers is approximately $10\%$, while for photon showers the average $\chi^2/\text{ndf}$ differs by less than $1\%$.

\begin{figure*}
    \centering
    \vspace{1em}
    \includegraphics[width=0.95\columnwidth]{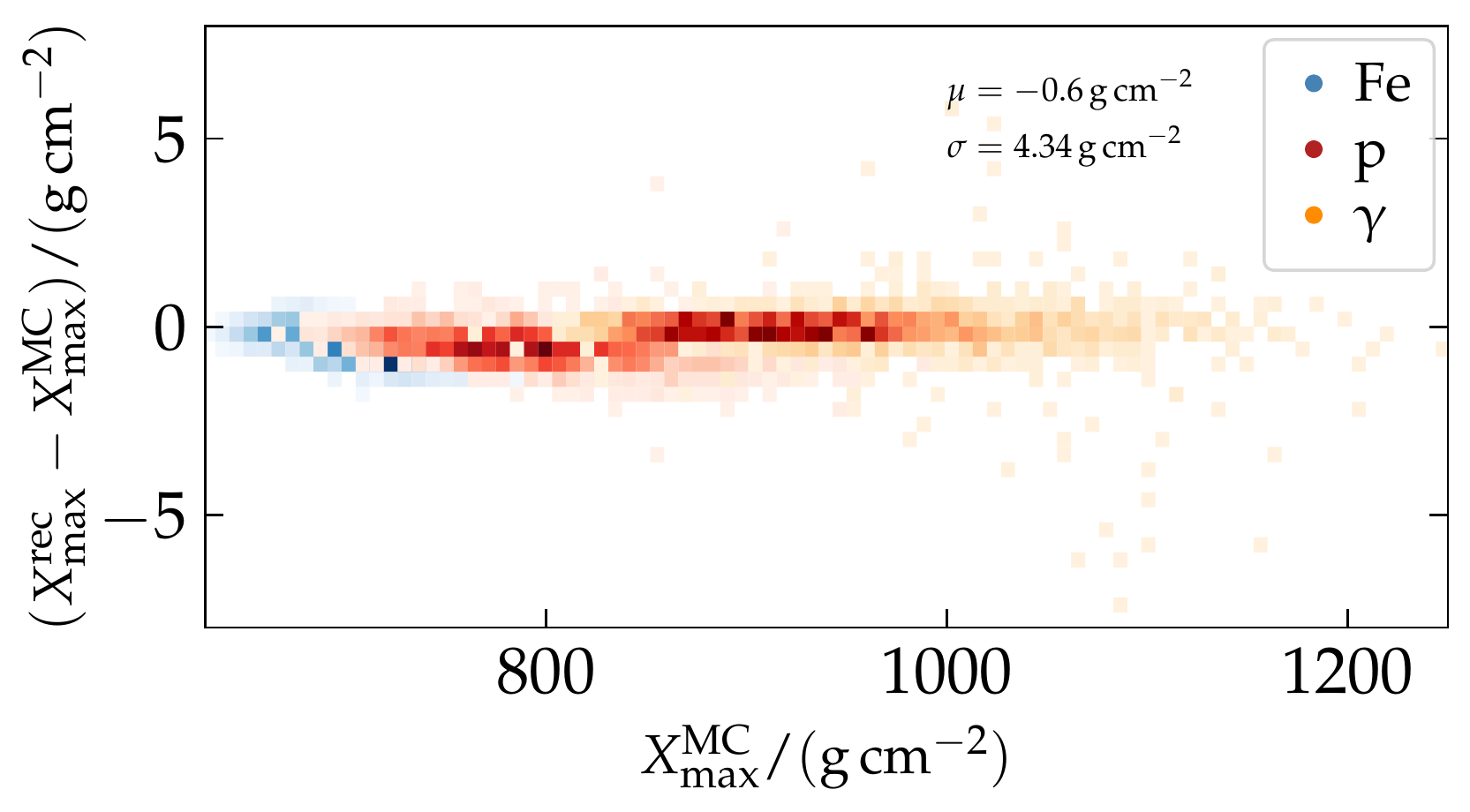}\hfill
    \includegraphics[width=0.95\columnwidth]{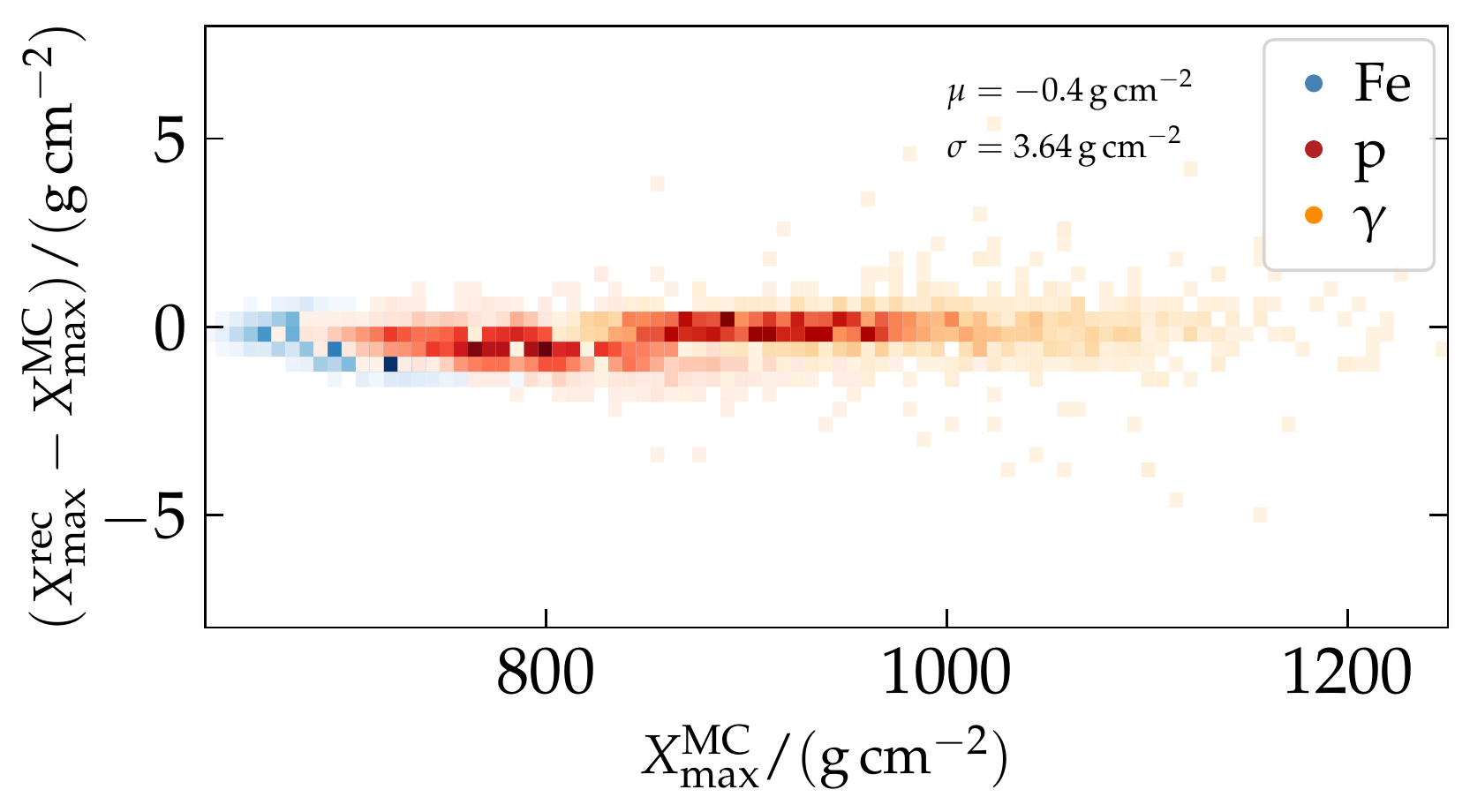}
    \caption{Distributions of the residuals of the reconstructed values of $\Xmax$ as a function of the \textsc{Conex} MC values of $\Xmax$ using the Greisen function (\emph{left}) and the GH function (\emph{right}).
    The showers were simulated with primary energies of $10^{18.5}\,\eV$, $10^{19}\,\eV$, and $10^{19.5}\,\eV$, using the \textsc{Sibyll2.3d} model of hadronic interactions.
    The overall mean and standard deviation of the distribution is given in the upper right corner.
    The distributions of residuals for the individual primary particles are colored accordingly.
    }
    \label{fig:xmaxbias}
\end{figure*}

Furthermore, we investigate the ability of the Greisen function to recover the depth of the shower maximum as well as the calorimetric energy deposit of the shower.
The distributions of the residuals $X_\text{max}^\text{rec} - X_\text{max}^\text{MC}$ as a function of the \textsc{Conex} MC values of $\Xmax$ are depicted in \cref{fig:xmaxbias}.
We observe that $\Xmax$ can be recovered very accurately from the simulated profile data using both functions, with an average precision of about $4\,\gcm$ for both fit functions.
The performance of the Greisen function of ``finding'' the right depth of the shower maximum is thus approximately equal as of the GH function.

\begin{figure*}
    \centering
    \vspace{1em}
    \includegraphics[width=0.95\columnwidth]{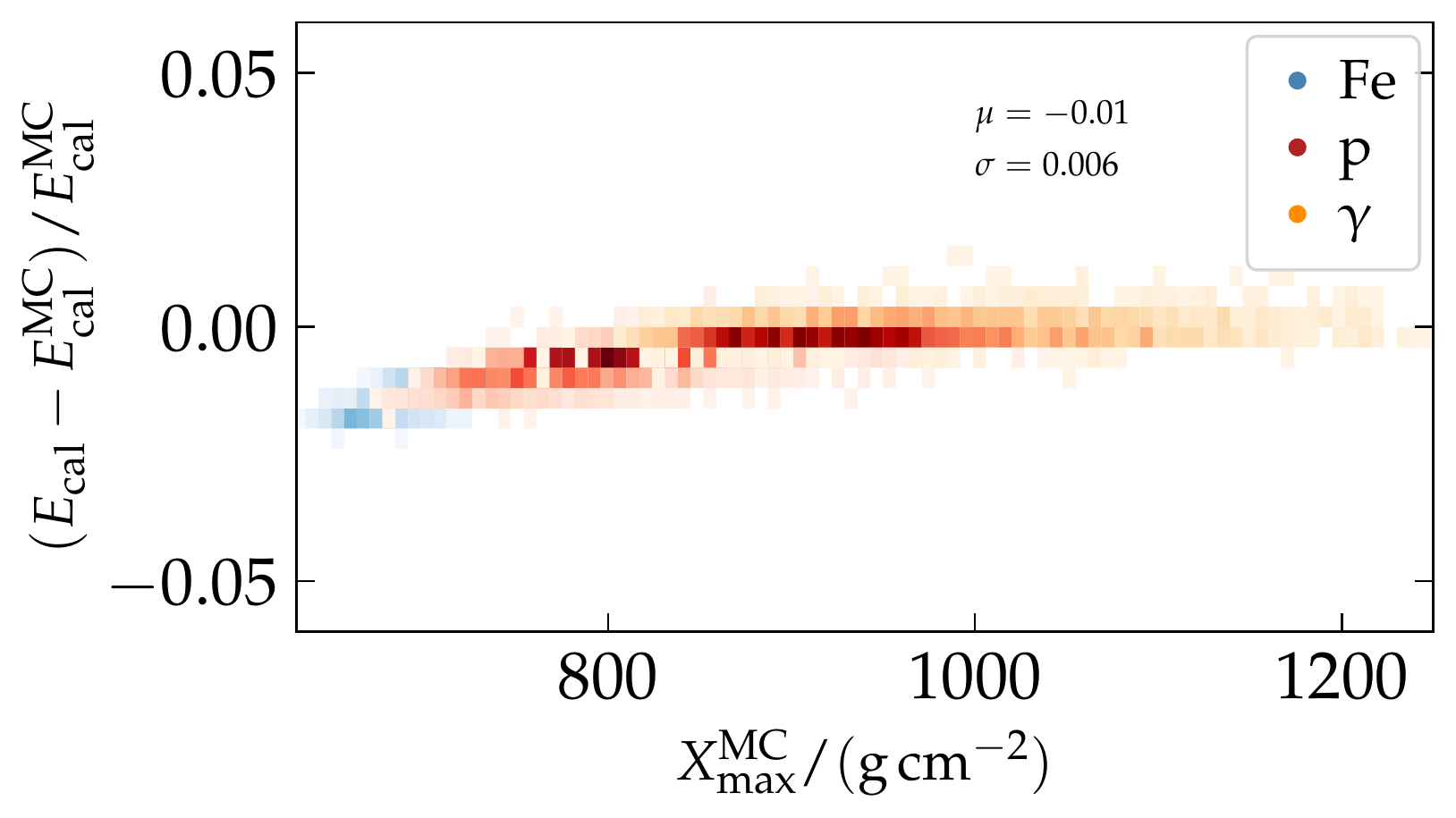}\hfill
    \includegraphics[width=0.95\columnwidth]{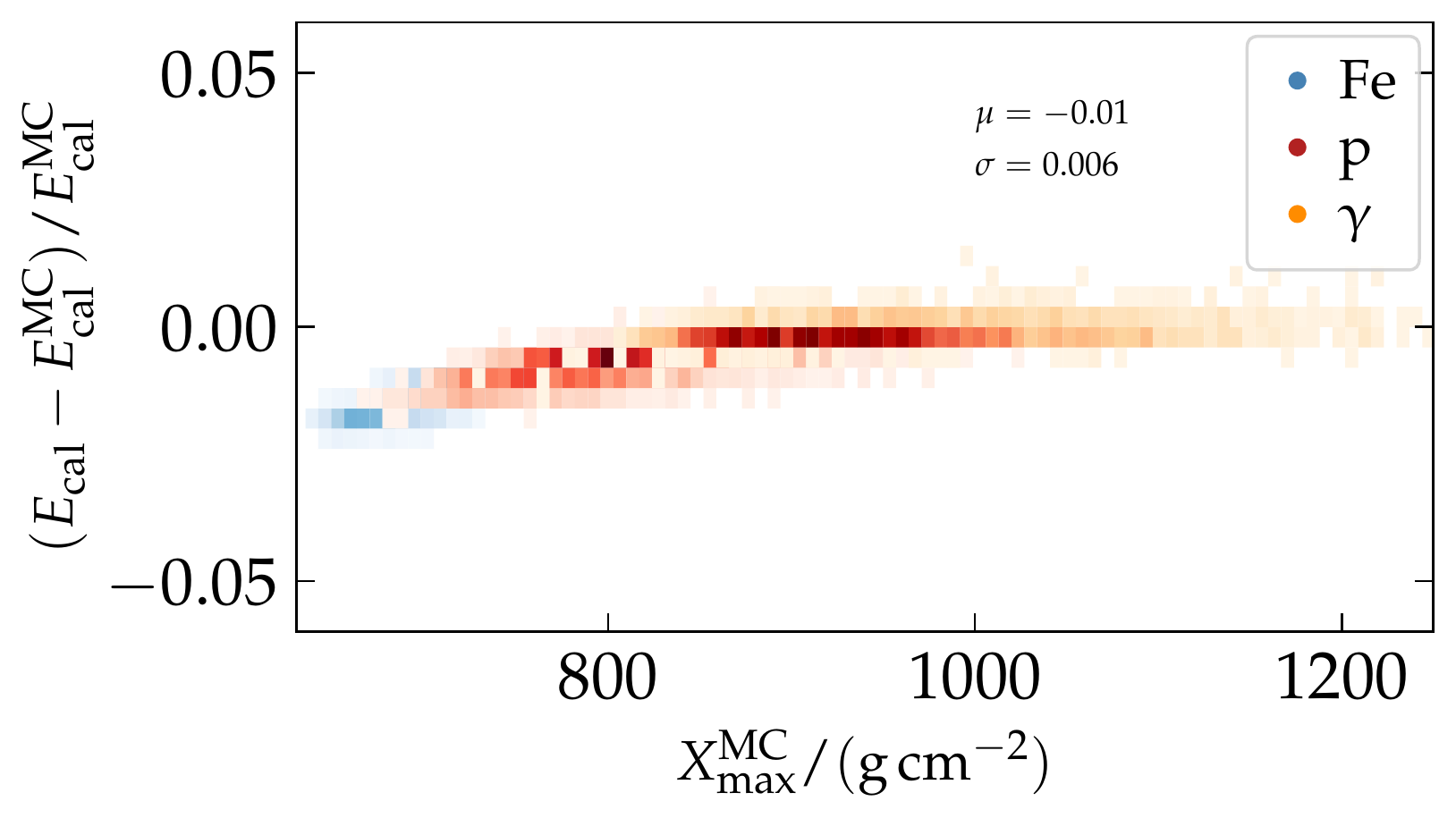}
    \caption{Distributions of the difference of the recovered calorimetric energy $E_\text{cal}$ and the simulated calorimetric energy deposit $E^\text{MC}_\text{cal}$ as a function of the \textsc{Conex} MC values of $\Xmax$ using the Greisen function (\emph{left}) and the GH function (\emph{right}).
    The showers were simulated with primary energies of $10^{18.5}\,\eV$, $10^{19}\,\eV$, and $10^{19.5}\,\eV$, using the \textsc{Sibyll2.3d} model of hadronic interactions.
    The overall mean and standard deviation of the distribution is given in the upper right corner.
    The distributions of residuals for the individual primary particles are colored accordingly.
    }
    \label{fig:ecalbias}
\end{figure*}

The calorimetric energy deposit of the shower can be obtained by integrating the fitted profile function from $X_1$ up to $\infty$.
To obtain the calorimetric energy deposit from the best-fit of both functions, we integrate numerically\footnote{The result for the calorimetric energy deposit as obtained from the fitted and numerically integrated Greisen function is approximately the same as using the formula given in \cref{eq:integrated}.} from the respective best-fit value of $X_1$ to $2000\,\gcm \lesssim \infty$.
The relative residuals of the recovered calorimetric energy deposit $E_\text{cal}$ with respect to the simulated calorimetric energy deposit $E^\text{MC}_\text{cal}$ are depicted in \cref{fig:ecalbias} as a function of the \textsc{Conex} MC values of $\Xmax$.
The accuracy and precision of the recovered calorimetric energy to estimate the primary energy is the same for the Greisen and the GH function for all primary particles (and all hadronic interaction models).
The performance of the Greisen function to estimate the primary energy of the particle initiating the shower is thus equal to the performance of the GH function.

Additionally, we present two-dimensional distributions of $X_1$ and $X_0$ obtained from the Greisen function fit, as well as $X_1$ and $\Lambda$ from the GH function fit in \cref{fig:x1lam}.
When fitting simulated data, both the Greisen and the GH function appear to have the same ``\textit{pathology}'' to produce mostly negative best-fit values for $X_1$.
Even though the best-fit values for $X_1$ obtained from the Greisen function are less negative than from the GH, this shows that the values obtained for $X_1$ from fitted data cannot be treated as the point of the first interaction.
Besides the absolute scale, the distributions for the best-fit values of $X_1$ and $X_0$ ($\Lambda$) depicted in \cref{fig:x1lam} appear to be very similar for both the Greisen and the GH function.

\section{Mass-composition sensitivity of the Greisen function}

In terms of the performance to obtain $\Xmax$ from simulated data, the Greisen function is not second to the GH function (cf.\ \cref{fig:xmaxbias}).
Studying the depth of the shower maximum, the same sensitivity to the mass composition as expected from the GH function can thus be achieved using the Greisen function.

\begin{figure}
    \centering
    \vspace{1em}
    \includegraphics[width=0.9\columnwidth]{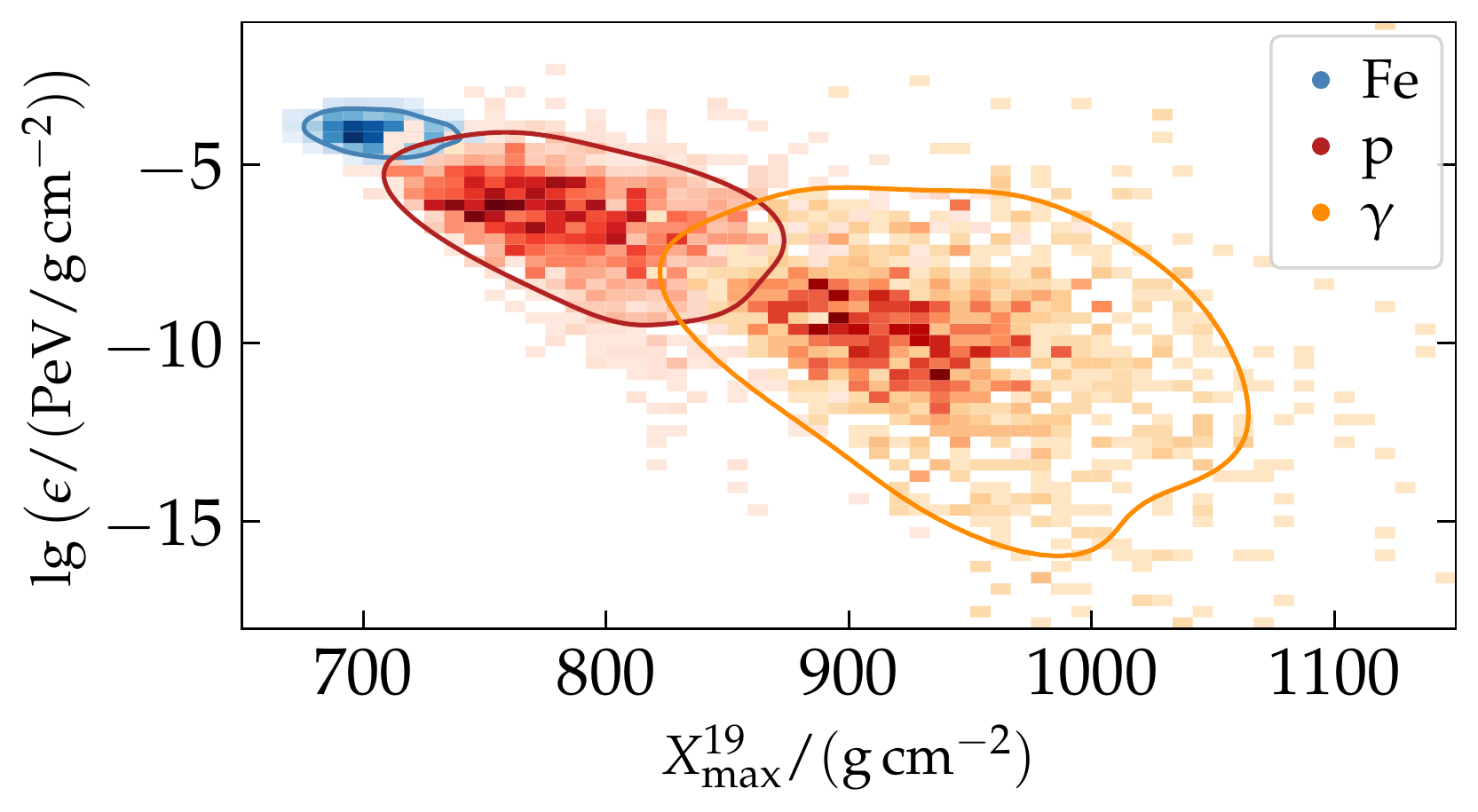}
    \caption{Two-dimensional distributions of the best-fit values of $\Xmax^{19}$ and $\epsilon$ using the Greisen function to fit simulated longitudinal profiles of showers with primary energies of $10^{18.5}\,\eV$, $10^{19}\,\eV$, and $10^{19.5}\,\eV$.
    The curved lines show the estimated $1\sigma$ extent of the respective distributions.
    All showers were simulated using the \textsc{Sibyll2.3d} model of hadronic interactions.
    }
    \label{fig:xmaxeps}
\end{figure}

Given the fact that $\epsilon$ is independent of the primary energy of the shower, and that the MC distributions of $\epsilon$ (cf.\ \cref{fig:calib_eps_ecut_values} (\emph{left})) are dependent on the type of primary particle, it is tempting to examine the primary-mass sensitivity of the best-fit values of $\epsilon$.
In \cref{fig:xmaxeps} we show the two-dimensional distributions of the best-fit results of $\Xmax$ and $\epsilon$.
To remove the direct dependence on the primary energy, instead of the true $\Xmax$, here we use
\begin{align}
    \Xmax^{19} := \Xmax - 58\,\gcm\,\lg\left(E_\text{cal}/10^{19}\,\eV\right),
\end{align}
assuming a constant decadal elongation rate of approximately $58\,\gcm$, and using $E_\text{cal}$ as obtained from the numerical integration of the best-fit function.

From \cref{fig:xmaxeps} it is obvious that the separation of the distributions of individual primary particles increases when the best-fit values of $\epsilon$ are considered alongside with $\Xmax$.
Numerically, the means of the distributions of $\Xmax^{19}$ obtained from a fit to proton and iron shower profiles are approximately 1.5 average standard deviations apart; on the diagonal line, which combines the information of $\epsilon$ and $\Xmax$, the means of the proton and iron distributions are separated by almost 1.9 average standard deviations\footnote{For the estimation, see \cref{app:fom}.}.
In case of photon-hadron separation, the distance improves from $1.4$ to $1.6$.
Thus, using \textsc{Conex} simulations, we see a clear improvement in terms of the separation of primary particles when employing the combination of $\epsilon$ and $\Xmax$ from the Greisen function over $\Xmax$ only.
Additional indicators for photon-like showers are the obtained values for $\chi^2$ and $X_0$ (cf.\ \cref{fig:chisq,fig:x1lam}). 
The Greisen function thus might be useful when trying to identify ultrahigh-energy photons in fluorescence detector data.

To compare against the behaviour of the GH function, in \cref{fig:xmaxnmax} we show the two-dimensional distributions of $\Xmax^{19}$ and $N_\text{max}^{19} = N_\text{max}/(E_\text{cal}/10^{19}\,\eV)$, as obtained from the GH function fitted to simulated air-shower data.
As can be seen from \cref{fig:xmaxnmax}, $N_\text{max}^{19}$ does not yield additional information about the mass of the primary particle on its own, while $\epsilon$ does.

\section{Fitting the Greisen function with fixed shape}
\label{sec:fixshape}

To boost the performance of the fitting procedure given only poor data, the GH function can be used with constraints to fix the shape of the function to an expected shape of the profile data \citep{PierreAuger:2019phh}.
These constraints are realized by reparametrizing the GH function in terms of ``$L$'' and ``$R$'' and then constraining these new parameters, which define the width and skewness of the function, respectively (see \cite{PierreAuger:2019phh} for details).
However, the average values of $L$ and $R$ depend on the primary particle of the shower \citep{Andringa:2011zz,Andringa:2012ju}.
Using the Greisen function, the shape of the profile can be determined simply by fixing or constraining the parameter $\epsilon$.
Moreover, one can easily choose whether the function should resemble an average gamma-ray, proton, or iron shower, depending on the corresponding value of $\epsilon$ (cf.\ \cref{fig:calib_eps_ecut_values}).

\begin{figure}
    \centering
    \vspace{1em}
    \includegraphics[width=0.9\columnwidth]{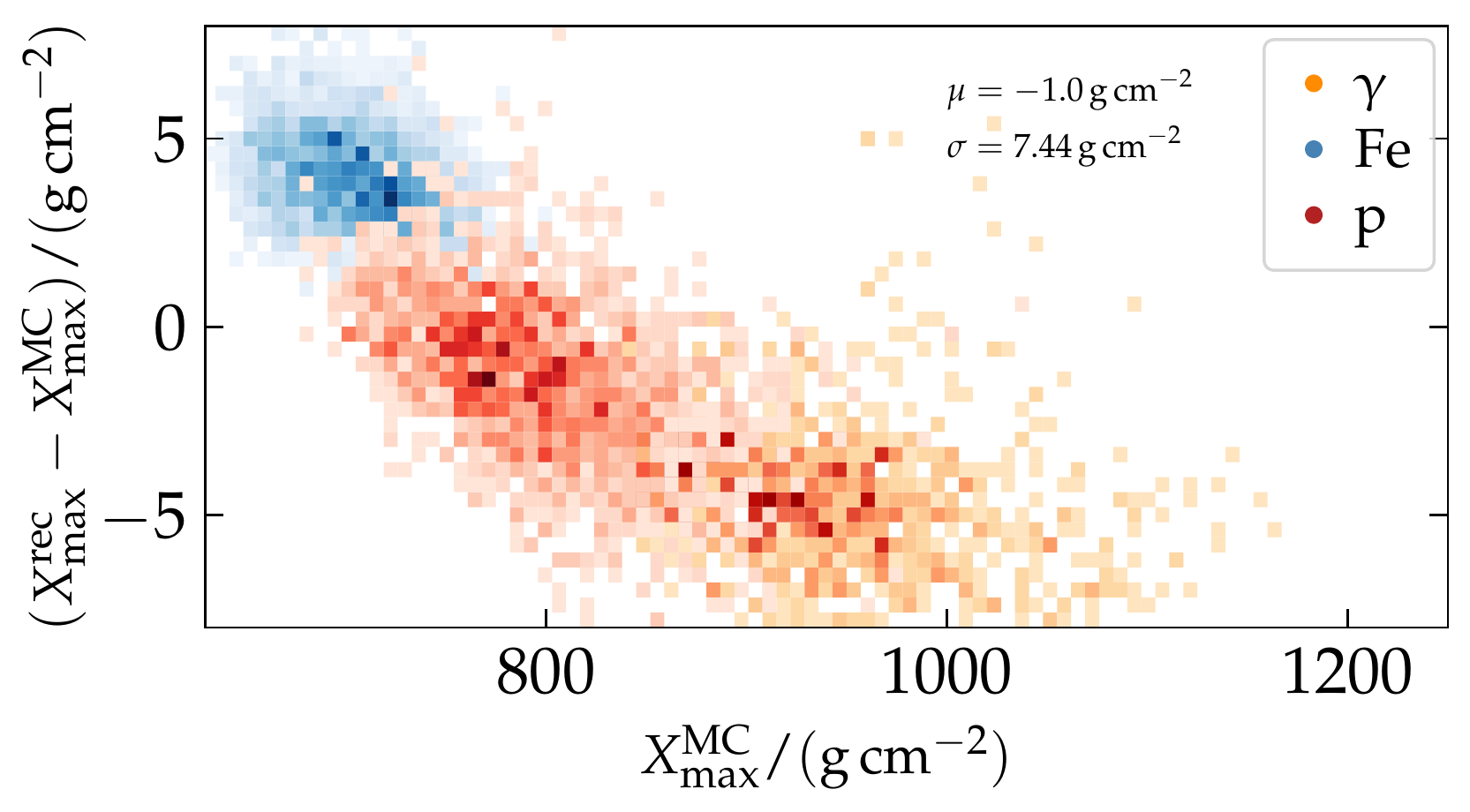}\\[0.5em]
    \includegraphics[width=0.9\columnwidth]{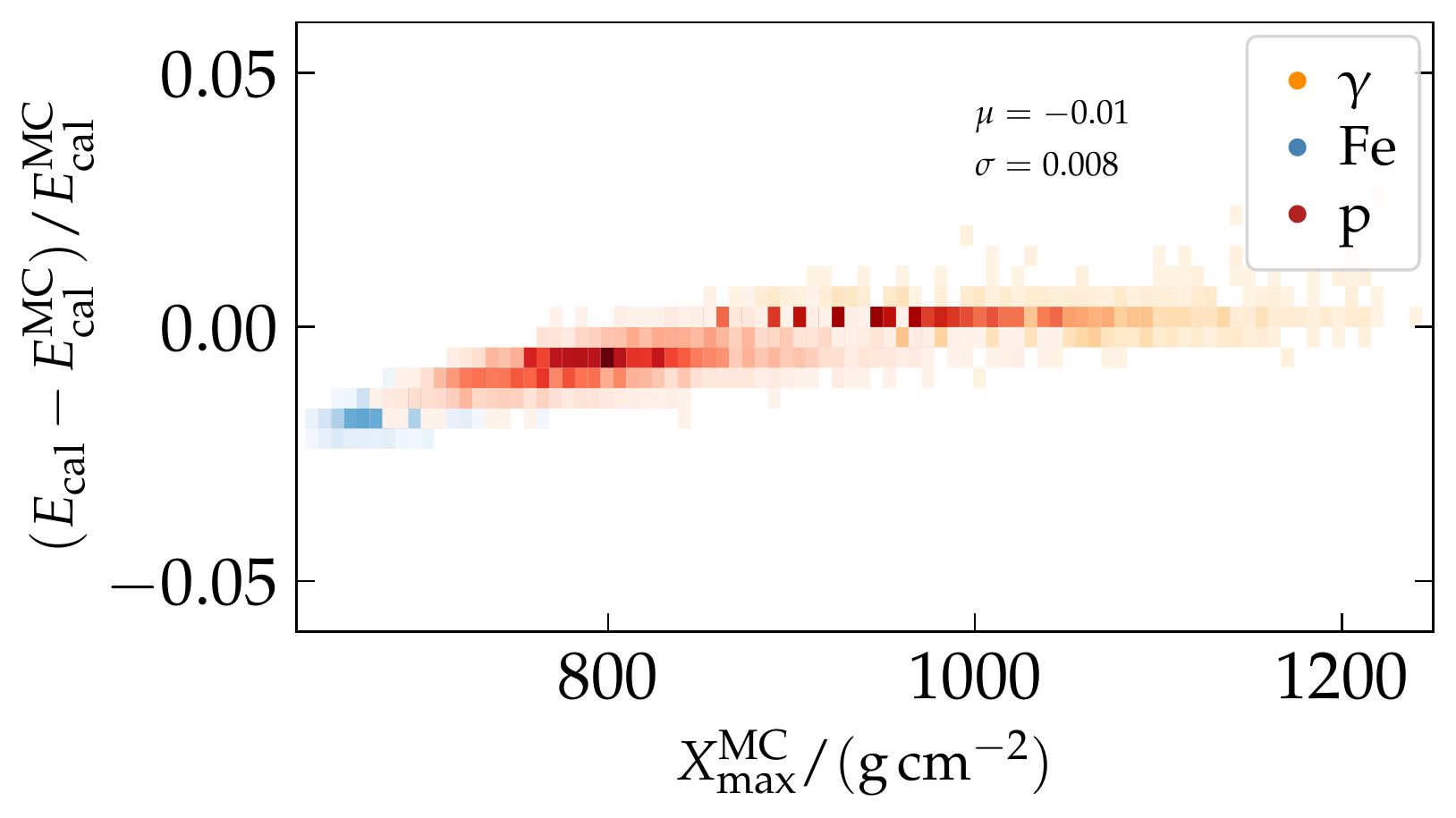}\\[0.5em]
    \includegraphics[width=0.9\columnwidth]{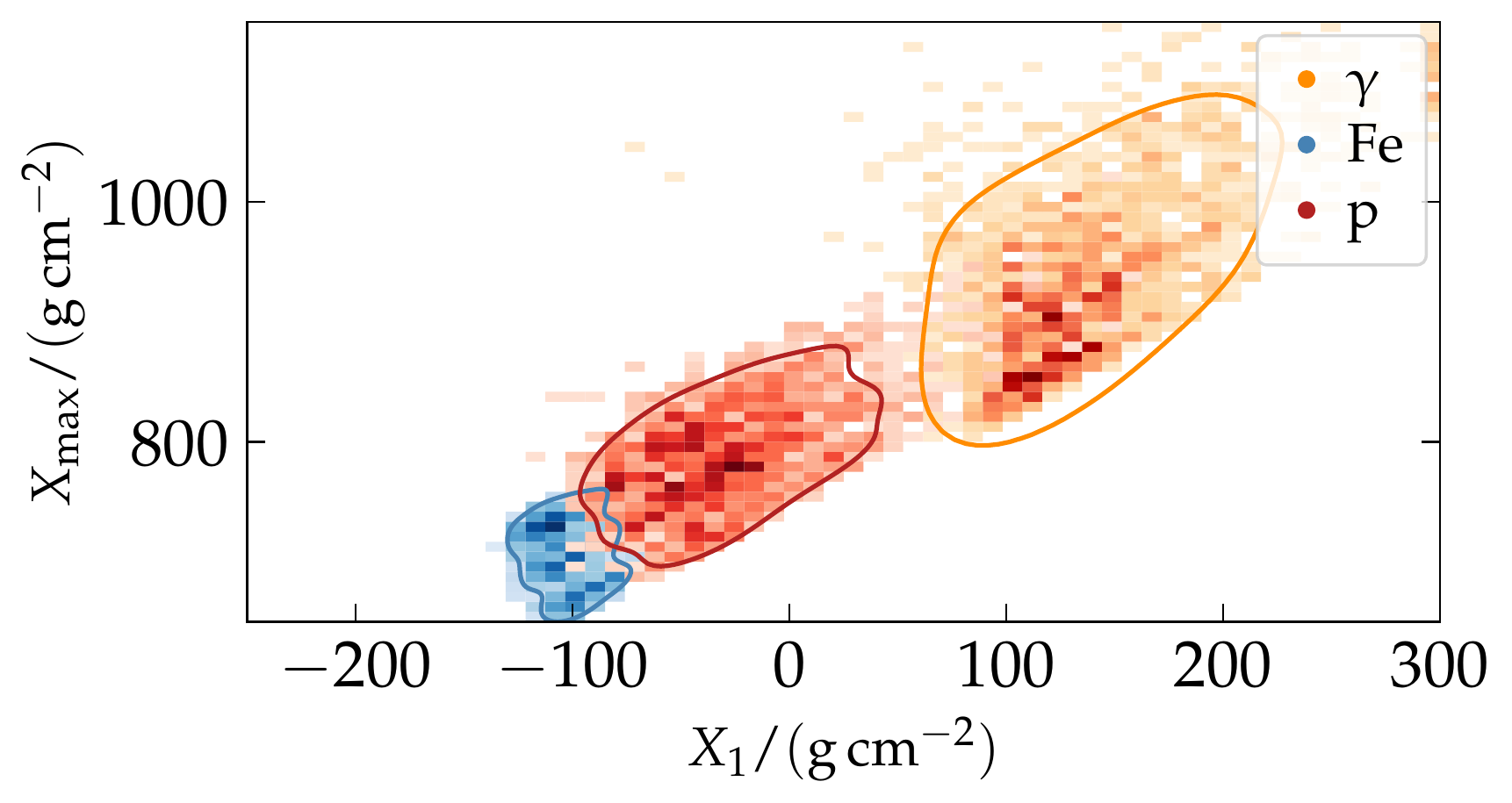}
    \caption{The $\Xmax$-bias (\emph{top}) and $E_\text{cal}$-bias (\emph{middle}) as a function of the \textsc{Conex} MC value of $\Xmax$, as well as the correlation of $\Xmax$ and $X_1$ using a fixed-shape Greisen function with only three free parameters to fit simulated air-shower data.
    The showers were simulated using the \textsc{Sibyll2.3d} model of hadronic interactions with primary energies of $10^{18.5}\,\eV$, $10^{19}\,\eV$, and $10^{19.5}\,\eV$.
    }
    \label{fig:fixed}
\end{figure}

To demonstrate, we fix $\epsilon = 10^{-6.2}\,\text{PeV}/(\gcm)$ as a compromise between iron-like and proton-like shower profiles and fit the simulated data with only three free parameters, namely $X_1$, $\Xmax$, and $X_0$.
As can be seen from \cref{fig:fixed}, the $\Xmax$ bias is minimal for proton showers, using the Greisen function with an hadron-like fixed shape, but increases for gamma-ray-induced showers and heavy nuclei.
The estimated calorimetric energy deposit, however, is almost unaffected (the bias for hadronic showers changes by $\approx0.5\%$), when using only three free parameters.
Lastly, the best-fit values of $\Xmax$ and $X_1$ become highly correlated for the Greisen function with fixed $\epsilon$, as it is expected from \cref{eq:beta2,eq:eps}.

\section{Discussion and Summary}

In this article, we present the Greisen function in its original form and discuss its relation to the shower age parameter $s$.
Furthermore, we present a way to derive the function from literature.
We show that with slight modifications the Greisen function can be rewritten to match individual simulated air-shower profiles, even from hadronic primaries.
We confirm this statement using simulated air showers from different primary particles at different energies, and using different hadronic interaction models.
Contrary to popular belief, the Greisen function matches the simulated air-shower profiles even somewhat better than the most commonly used Gaisser--Hillas profile function.
In contrary to the Gaisser--Hillas function, which was introduced as an alternative to the Constant Intensity Cut method, the Greisen function was derived to describe the full longitudinal profiles of air showers.

We analyse the performance of the Greisen function to recover air-shower observables from simulated profile data assuming an ideal detector.
In this analysis, we show that the Greisen function yields approximately the same performance as the Gaisser--Hillas function to determine the calorimetric energy deposit and the depths of the shower maxima from simulated showers at different primary energies using different hadronic interaction models.

We identified the shape-parameter $\epsilon$ of the Greisen function, which is primary-mass sensitive and can help distinguishing different types of primary particles, additionally to the slanted depth of the shower maximum $\Xmax$.
Lastly, we demonstrate that fixing $\epsilon$ can elegantly fix the shape of the Greisen function and thus shower profiles can be fitted even with only three free parameters.
In this case, while the recovered values of $\Xmax$ are slightly biased, the accuracy and precision of the obtained calorimetric energy deposit are unaffected.

We conclude that the Greisen function proves itself useful to describe air-shower profiles and bears additional potential for photon-hadron separation as well as mass-composition studies, and could thus be used in the search for light particles in air-shower data.

\section{Acknowledgements}

The authors would like to thank Alexey Yushkov, Eva Santos, Armando di Matteo, and Darko Veberi\v{c} for fruitful comments and discussion as well as the referee for valuable comments and critique.
This work was partially supported by the Ministry of Education Youth and Sports of the Czech Republic and by the European Union under the grant FZU researchers, technical and administrative staff mobility, registration number CZ.02.2.69/0.0/0.0/18\_053/0016627. This work was supported by the Czech Science Foundation – Grant No. 21-02226M.

\section{Code Availability}

The analysis code for this article is available upon request under:
\begin{verbatim}gitlab.com/stadelmaier/greisen_fct_fit\end{verbatim}

\clearpage 

\bibliography{bib}{}

\appendix
\begin{widetext}

\section{Additional Expressions}

\subsubsection*{Alternative Form of the Greisen Function}
The Greisen function can be rewritten as

\begin{align}
    N(t) = \frac{\epsilon}{\sqrt{\beta}} \, \e^{t^\prime\left(1-\tfrac{3}{2}\ln s\right)} = \frac{\epsilon}{\sqrt{\beta}}  \left( \frac{3 t^\prime}{t^\prime + 2 \beta} \right)^{-\tfrac{3}{2} t^\prime}\,\e^{t^\prime},
    \label{eq:greisen_alternative}
\end{align}
with $t^\prime = t - t_1$ and $\beta = t_\text{max} - t_1$.

\subsubsection*{Integral of the Greisen Function}
Numerically, we find that the integrated Greisen profile, which yields the calorimetric energy deposit and is thus a good estimator for the primary energy, can be approximated within ${\sim}0.5\%$ by

\begin{align}
E_\text{cal} = \int\limits_{X_1}^\infty \frac{\epsilon}{\sqrt{\beta}} \, \exp\left[(t-t_1)\left(1-\tfrac{3}{2}\ln s\right)\right]\dd X \simeq 3.1\,\epsilon\, \e^\beta\, X_0,
\label{eq:integrated}
\end{align}
where identified the pre-factor of ${\sim}3.1$ numerically as suitable for different values of $X_0$ and $X_1$.
Thus, we find that 
\begin{align}
    \epsilon \simeq \frac{E_\text{cut}}{3.1\, X_0},
\end{align}
which is in good agreement with \cref{eq:eps_ecut_relation}.

\subsubsection*{Proton-Iron Separation Using the Greisen Function}

The mean and standard devations of the distributions of $\Xmax^{19}$ as well as of $\lg(\epsilon/(\text{PeV}/(\gcm)))$ depicted in \cref{fig:xmaxeps} are given in \cref{tab:dists}.
In units of the mean standard deviation of the distributions, the average values of $\Xmax^{19}$ of protons and iron nuclei induced showers are separated by
\begin{align}
    \xi = \frac{\vert\mu_\text{p}(\Xmax^{19}) - \mu_\text{Fe}(\Xmax^{19})\vert}{\sqrt{\sigma_\text{p}(\Xmax^{19})^2 + \sigma_\text{Fe}(\Xmax^{19})^2}} \simeq 1.52.
\end{align}
For $\hat{\epsilon} = \lg(\epsilon/(\text{PeV}/(\gcm)))$, this distance evaluates to 
\begin{align}
    \zeta = \frac{\vert \mu_\text{p}(\hat{\epsilon}) - \mu_\text{Fe}(\hat{\epsilon})\vert}{\sqrt{\sigma_\text{p}(\hat{\epsilon})^2 + \sigma_\text{Fe}(\hat{\epsilon})^2}} \simeq 1.18.
\end{align}
Assuming no covariance between $\hat{\epsilon}$ and $\Xmax^{19}$, we find
\begin{align}
    \sqrt{\xi^2+\zeta^2} = 1.93.
    \label{app:fom}
\end{align}
Note that \cref{app:fom} is a generous estimate, since it is clear from \cref{fig:xmaxeps} that the covariance between $\hat{\epsilon}$ and $\Xmax^{19}$ is not zero.

\begin{table}[]
    \centering
    \caption{Mean $\mu$ and standard deviations $\sigma$ of the distributions of $\Xmax^{19}$ and $\epsilon$ as depicted in \cref{fig:xmaxeps}.}
    \begin{tabular}{c c c}
      ($\mu$, $\sigma$)   & $\Xmax^{19}/(\gcm)$ & $\lg(\epsilon/(\text{PeV}/(\gcm)))$ \\
         \hline
       $\upgamma$  & (984.4, 107.1) & (-11.312, 4.33) \\
       p & (812.0, 62.2) & (-7.217, 2.57) \\
       Fe & (711.0, 22.5) & (-4.124, 0.49) 
    \end{tabular}
    \label{tab:dists}
\end{table}

\section{Additional Figures}
\label{app:clib}

We show the effect of shower-to-shower fluctuations for showers simulated with different hadronic interaction models and at different primary energies. 
\cref{fig:calib_all_hims} depicts the universal relation between $\epsilon$ and $E_\text{cut}$ even for the \textsc{Epos-LHC} and \textsc{QgsjetII-04} hadronic interaction models, using different primary particles and different primary energies.
In \cref{fig:calib_eps_ecut_values} we present the individual distributions of $\epsilon$ and $E_\text{cut}$ for all three hadronic interaction models.
Note that the histograms of the distributions for photon showers in \cref{fig:calib_eps_ecut_values} are slightly truncated, as for individual showers $\epsilon$ ($E_\text{cut}$) reaches values down to ${\approx}1\,\text{keV}/\gcm$ (${\approx}1\,\MeV$) (cf.\ \cref{fig:ecut_eps_corr,fig:calib_all_hims}).

\begin{figure}
    \centering
    \vspace{1em}
    \includegraphics[width=0.45\columnwidth]{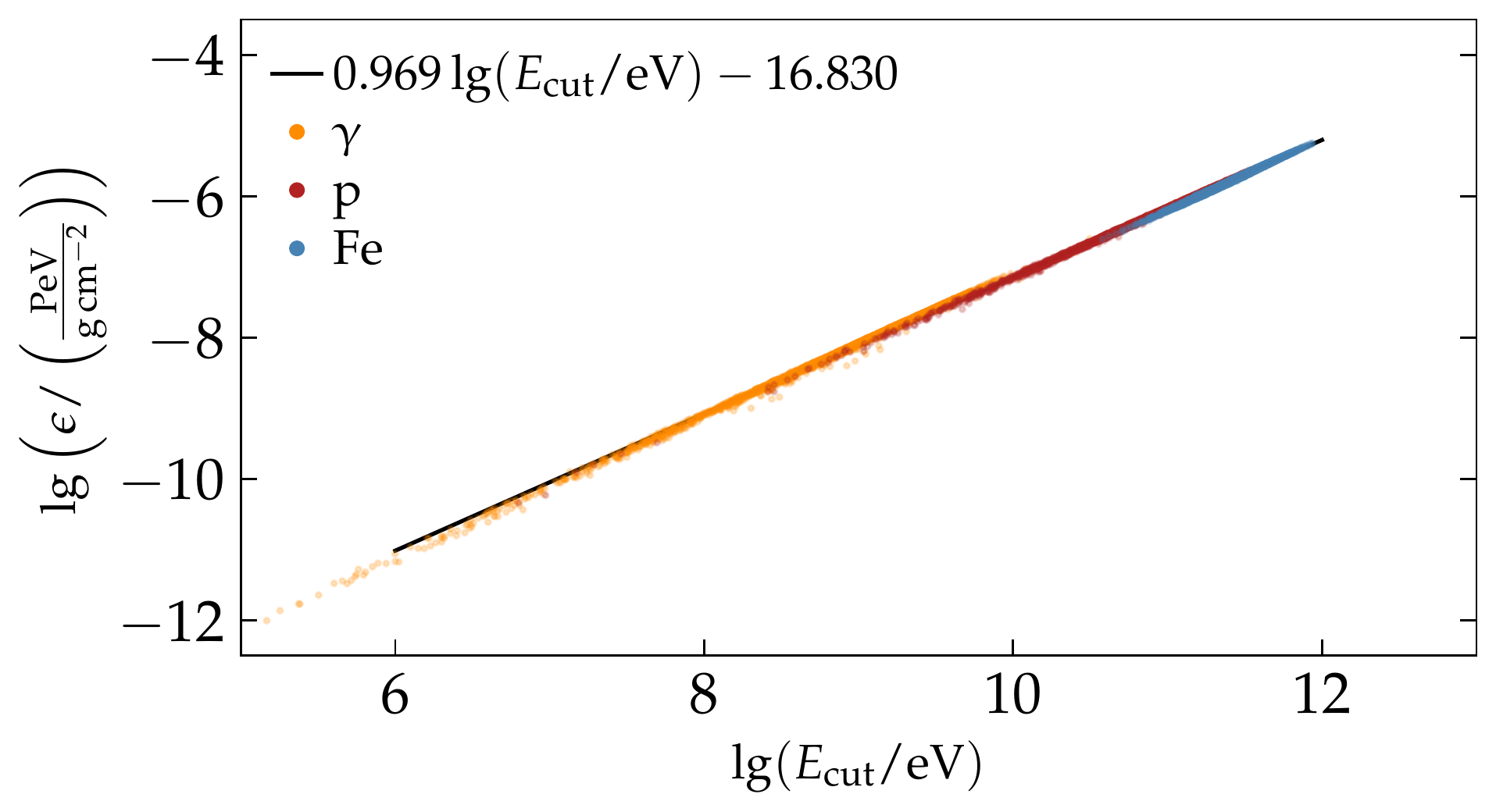}\hfill
    \includegraphics[width=0.45\columnwidth]{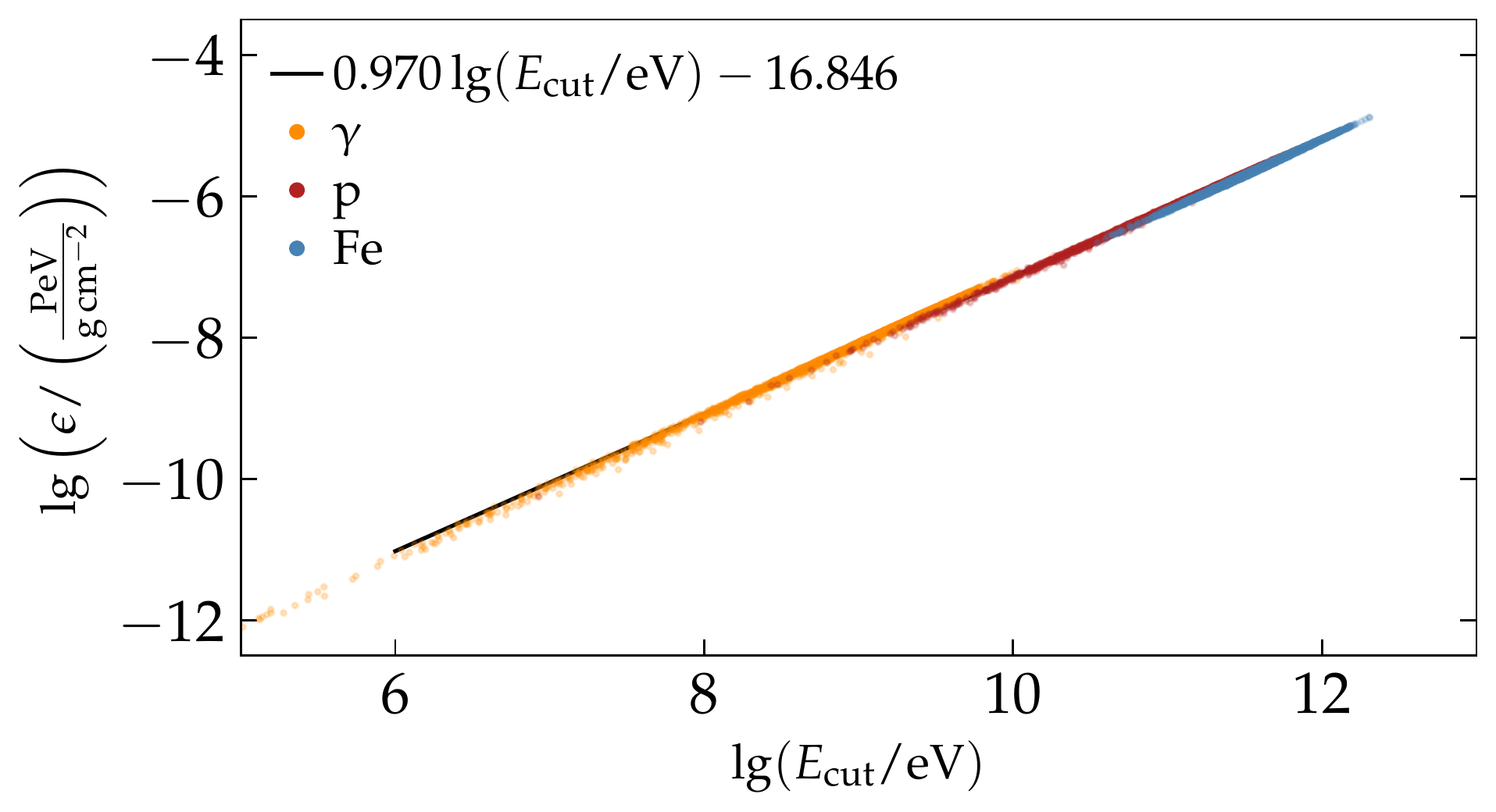}
    \caption{Relation of $\epsilon$ and $E_\text{cut}$ for simulated iron (blue), proton (red), and gamma-ray (orange) showers using the \textsc{Epos-lhc} (\emph{left}) and \textsc{QgsjetII-04} (\emph{right}) model of hadronic interactions.
    $E_\text{cut}$ and $\epsilon$ were obtained using \cref{eq:ecut} and \cref{eq:eps}, respectively. 
    In each panel we show results from 9000 showers using gamma-rays, protons, and iron nuclei primary particles with simulated energies of $10^{18.5}\,\eV$, $10^{19.0}\,\eV$, and $10^{19.5}\,\eV$ (1000 for each configuration).}
    \label{fig:calib_all_hims}
\end{figure}

\begin{figure}
    \centering
    \vspace{1em}
    \includegraphics[width=0.45\columnwidth]{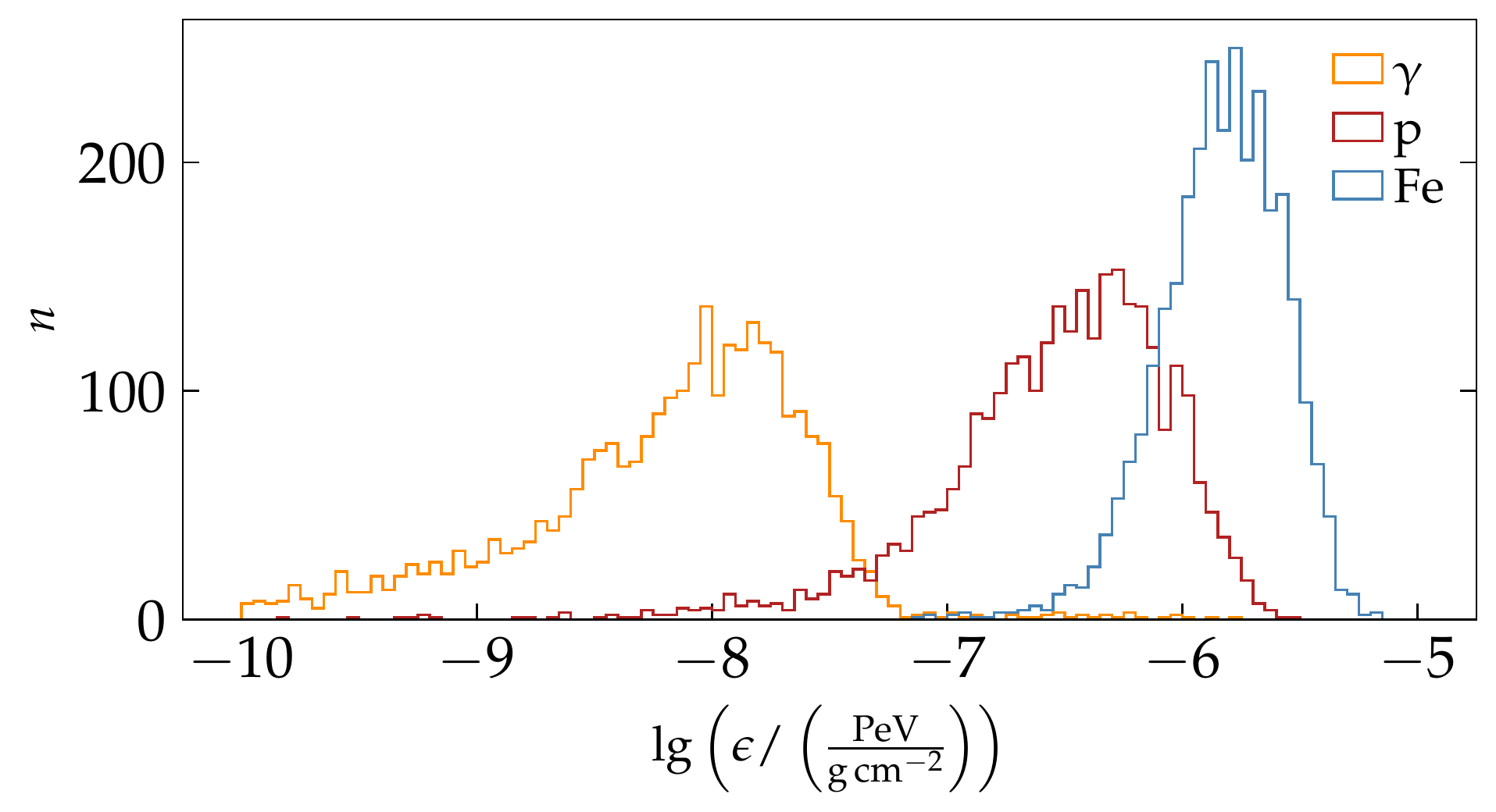}\hfill
    \includegraphics[width=0.45\columnwidth]{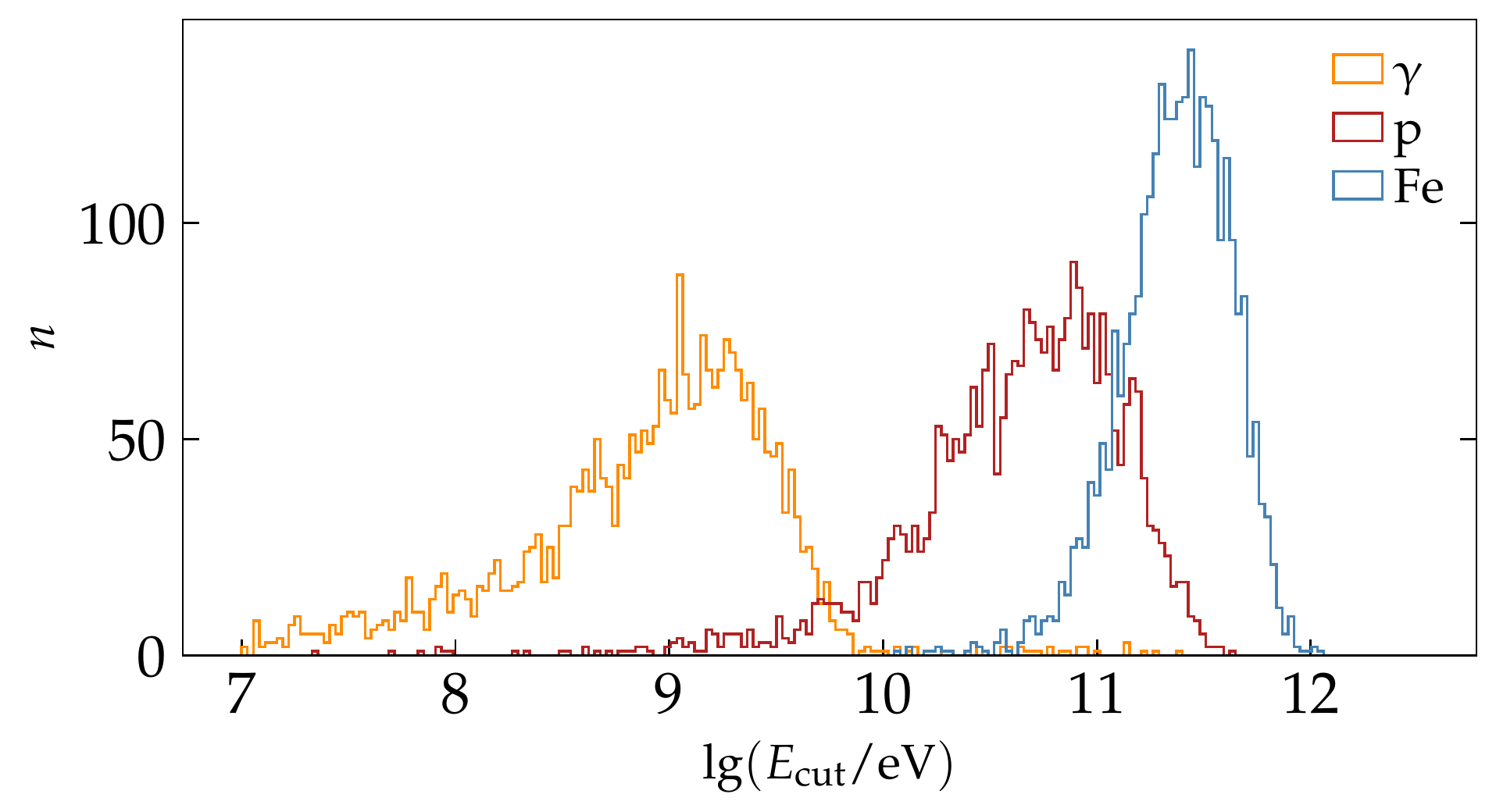} \\[0.5em]
    \includegraphics[width=0.45\columnwidth]{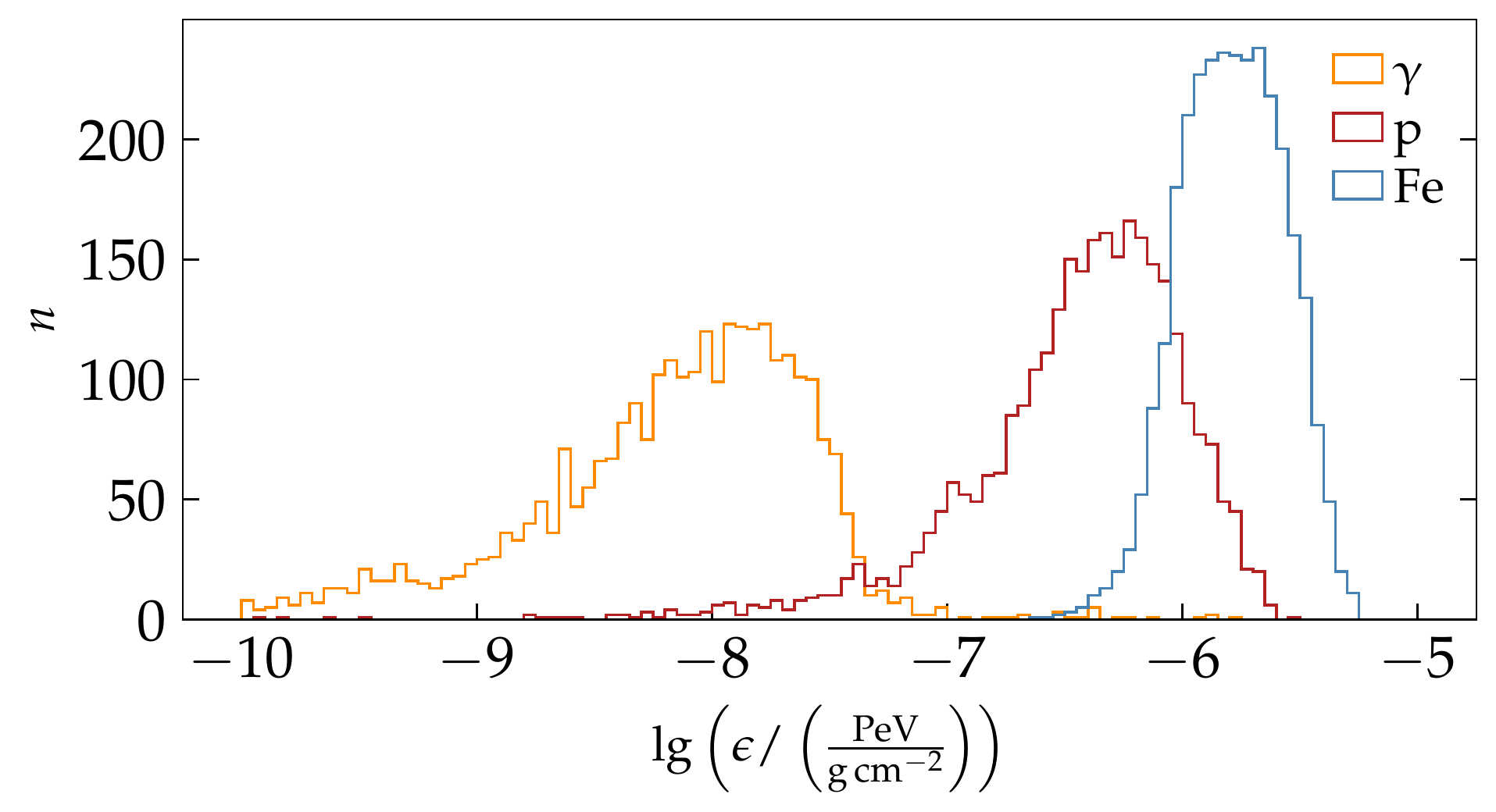}\hfill
    \includegraphics[width=0.45\columnwidth]{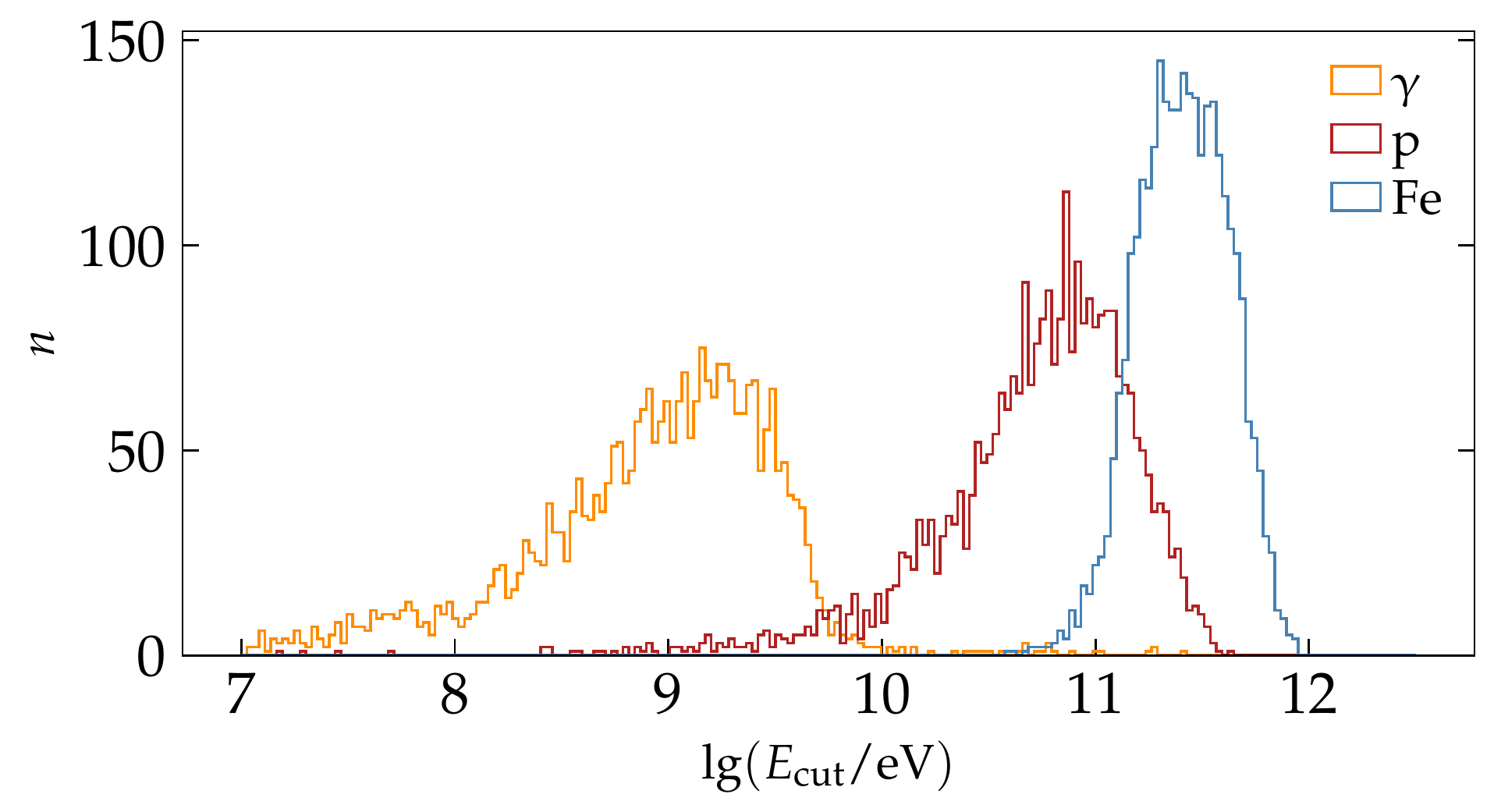} \\[0.5em]
    \includegraphics[width=0.45\columnwidth]{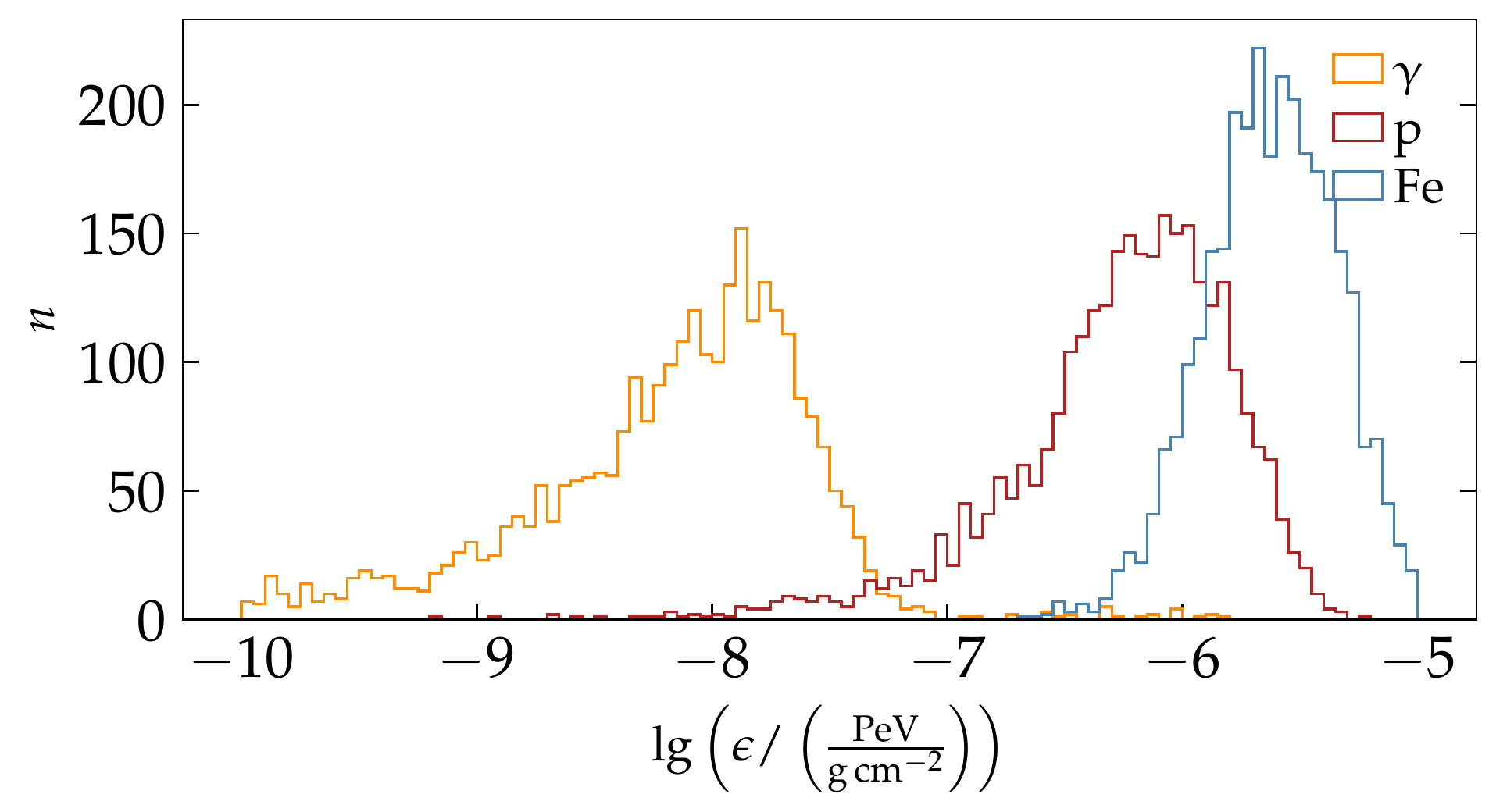}\hfill
    \includegraphics[width=0.45\columnwidth]{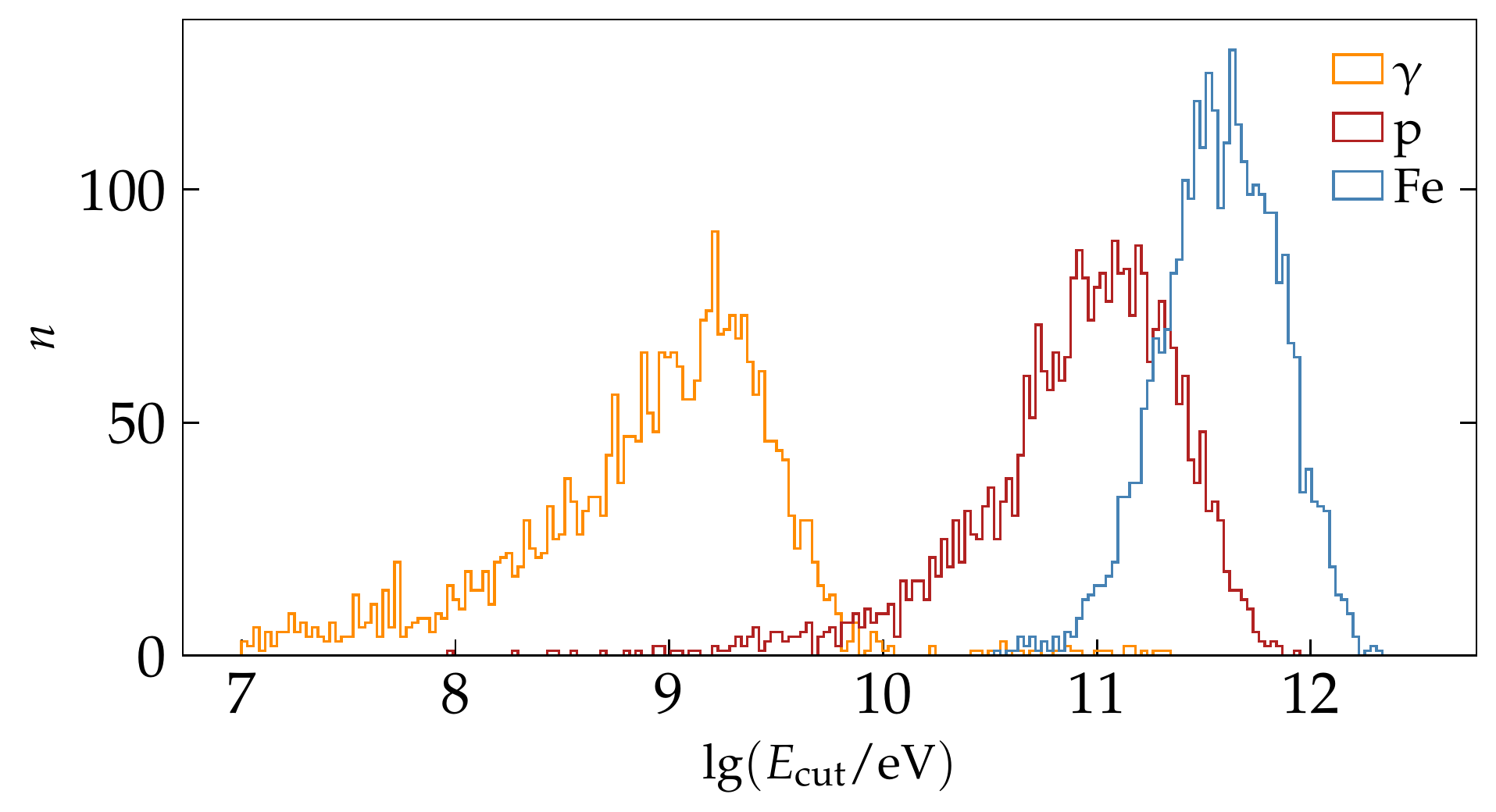} \\[0.5em]
    \caption{Distributions of $\epsilon$ (\emph{left}) and $E_\text{cut}$ (\emph{right}) calculated according to \cref{eq:ecut,eq:eps} for showers simulated using the (\emph{top to bottom}) \textsc{Sibyll2.3d}, \textsc{Epos-LHC}, and the \textsc{QgsjetII-04} models of hadronic interactions.
    In each panel we show results from 9000 showers using gammay-rays, protons, and iron nuclei as primary particles with simulated energies of $10^{18.5}\,\eV$, $10^{19.0}\,\eV$, and $10^{19.5}\,\eV$ (1000 for each configuration).}
    \label{fig:calib_eps_ecut_values}
\end{figure}

\begin{figure}
    \centering
    \vspace{1em}
    \includegraphics[width=0.45\columnwidth]{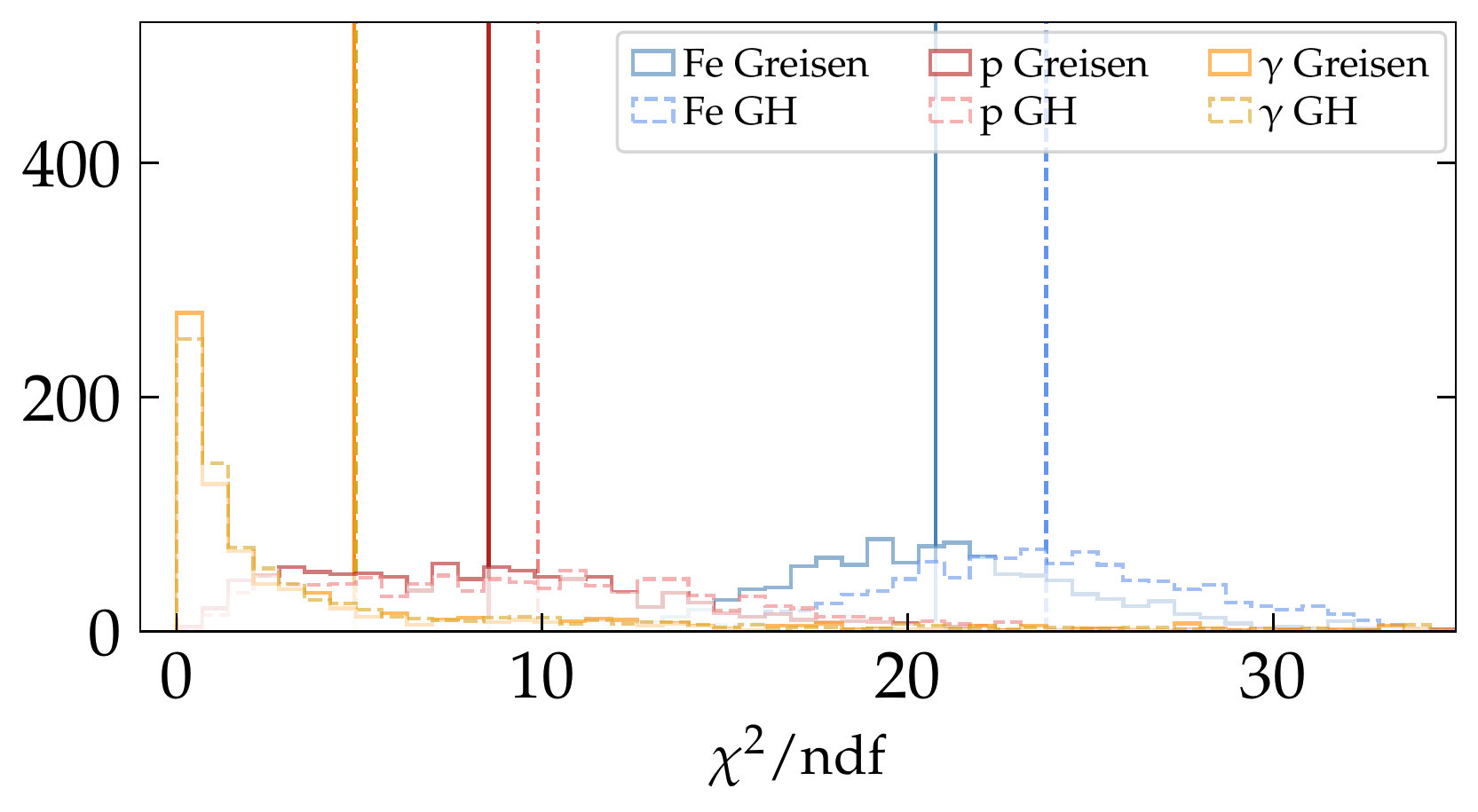}\hfill
    \includegraphics[width=0.45\columnwidth]{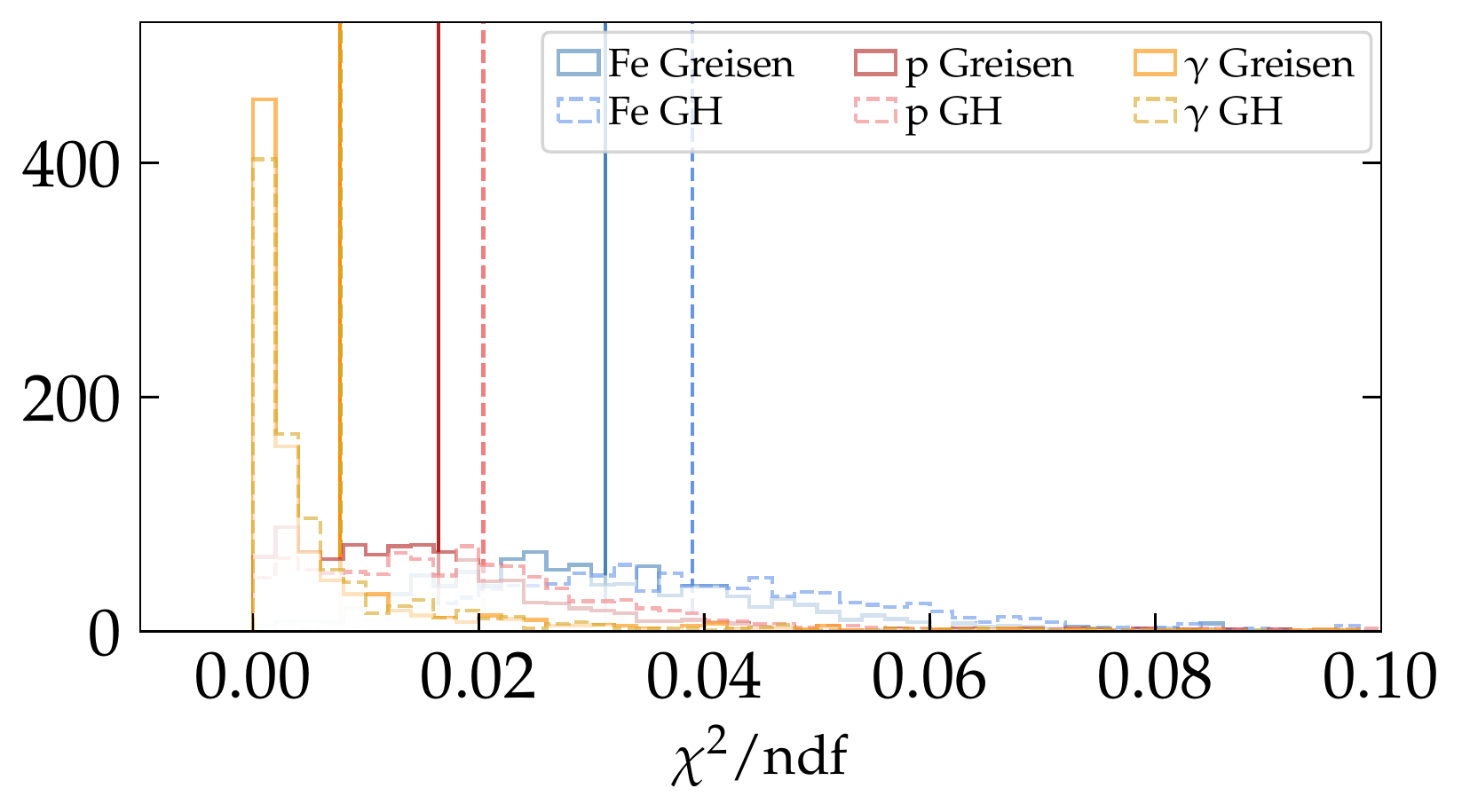}
    \caption{$\chi^2$-distributions for the Greisen and the GH functions fitted to simulated \textsc{Epos-LHC} air showers with primary energies of $10^{19.5}\,\eV$ (\emph{left}), and fitted to simulated \textsc{QgsjetII-04} air showers with primary energies of $10^{18.5}\,\eV$ (\emph{right}).
    The distributions for the Greisen (GH) functions are shown as a full (dashed) line, for each of the three primary particles.
    The respective mean of each distribution is indicated by a vertical line.
    }
    \label{fig:chisq_additional}
\end{figure}

\begin{figure}
    \centering
    \vspace{1em}
    \includegraphics[width=0.45\columnwidth]{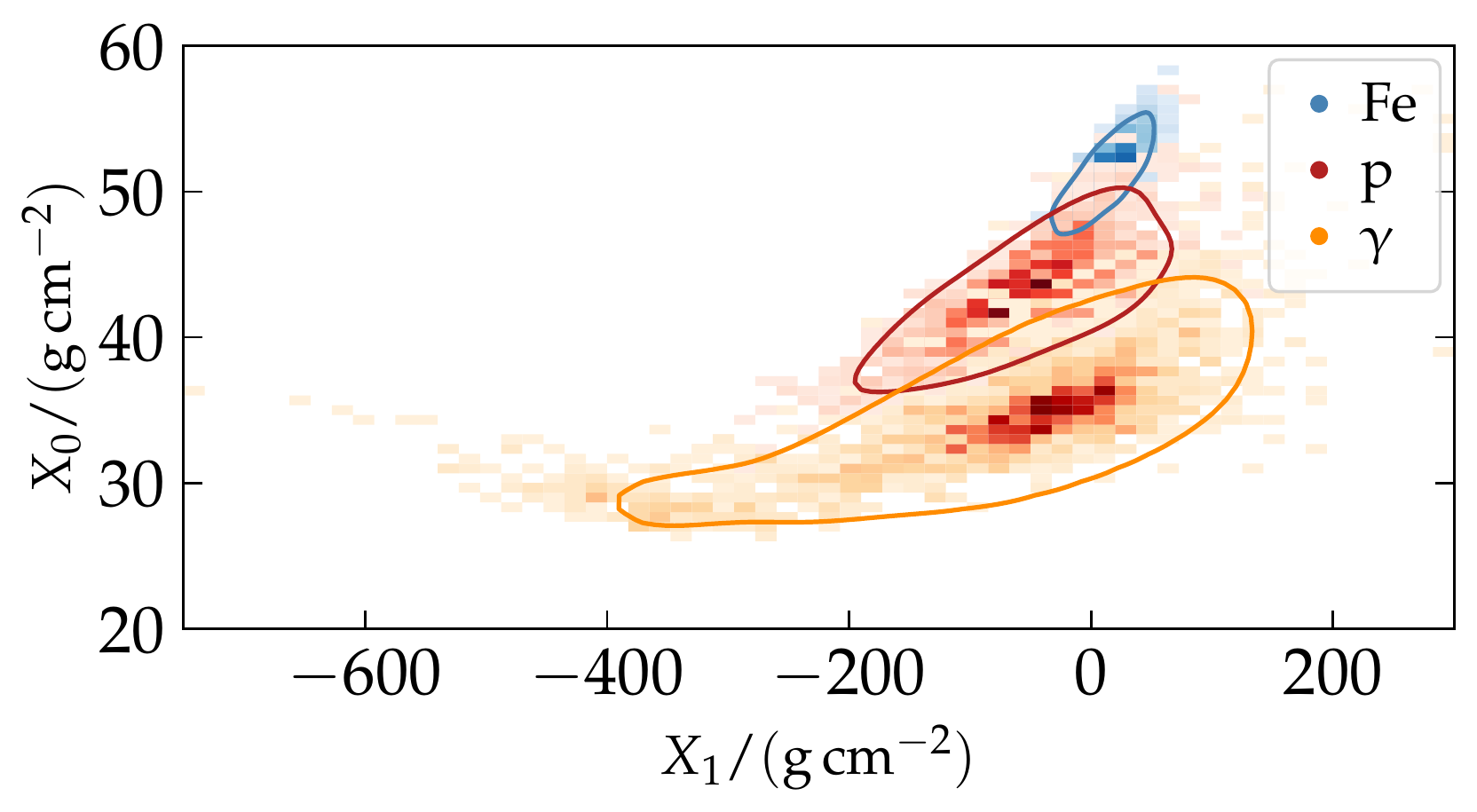}\hfill
    \includegraphics[width=0.45\columnwidth]{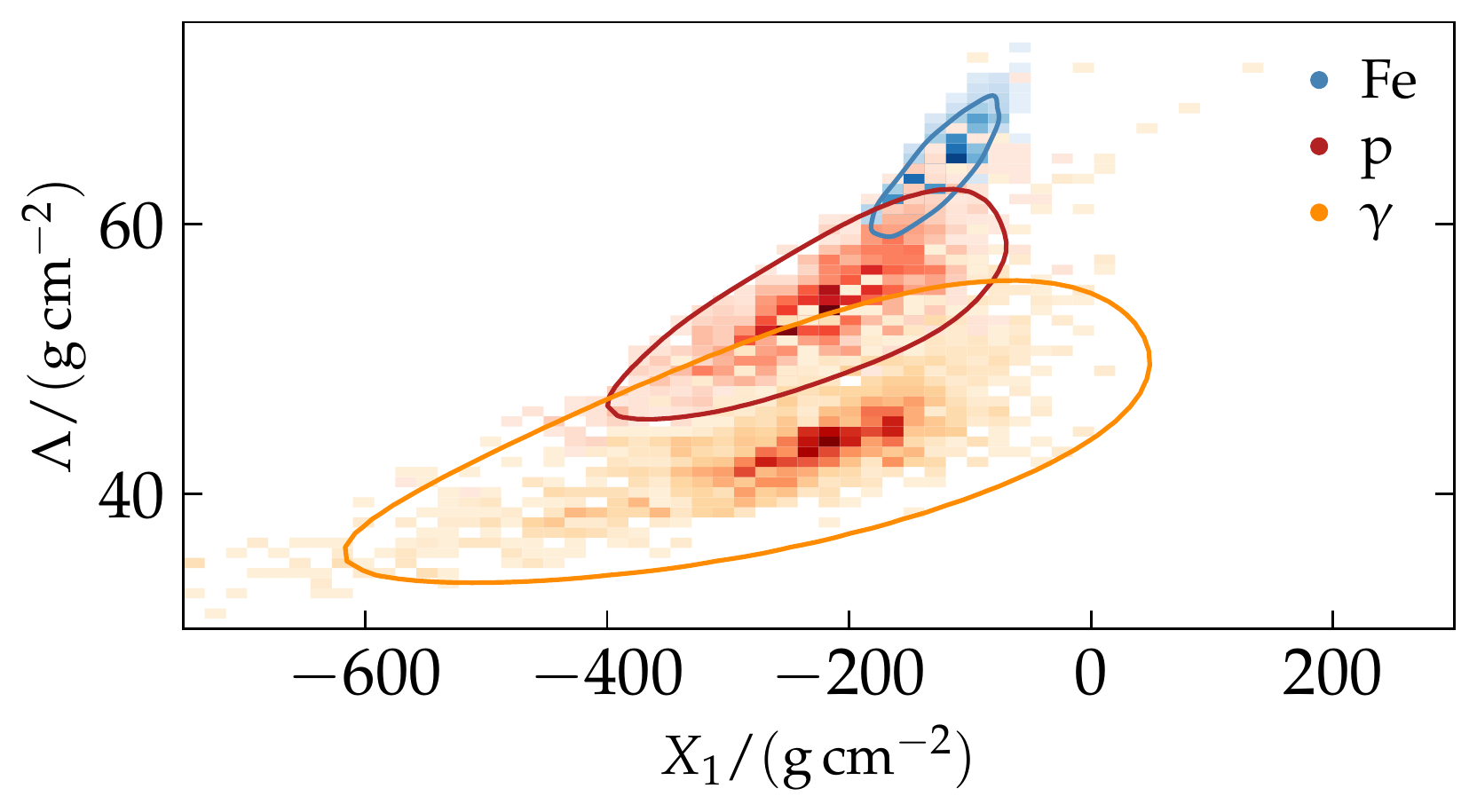}
    \caption{Two-dimensional distribution of the best-fit values of $X_1$ and $X_0$ using the Greisen function to fit simulated longitudinal profiles (\emph{left}), and $X_1$ and $\Lambda$ using the GH function (\emph{right}). The showers were simulated with primary energies of $10^{18.5}\,\eV$, $10^{19}\,\eV$, and $10^{19.5}\,\eV$, using the \textsc{Sibyll2.3d} model of hadronic interactions.
    The curved lines show the estimated $1\sigma$ extent of the respective distributions.
    }
    \label{fig:x1lam}
\end{figure}

\begin{figure}
    \centering
    \vspace{1em}
    \includegraphics[width=0.45\columnwidth]{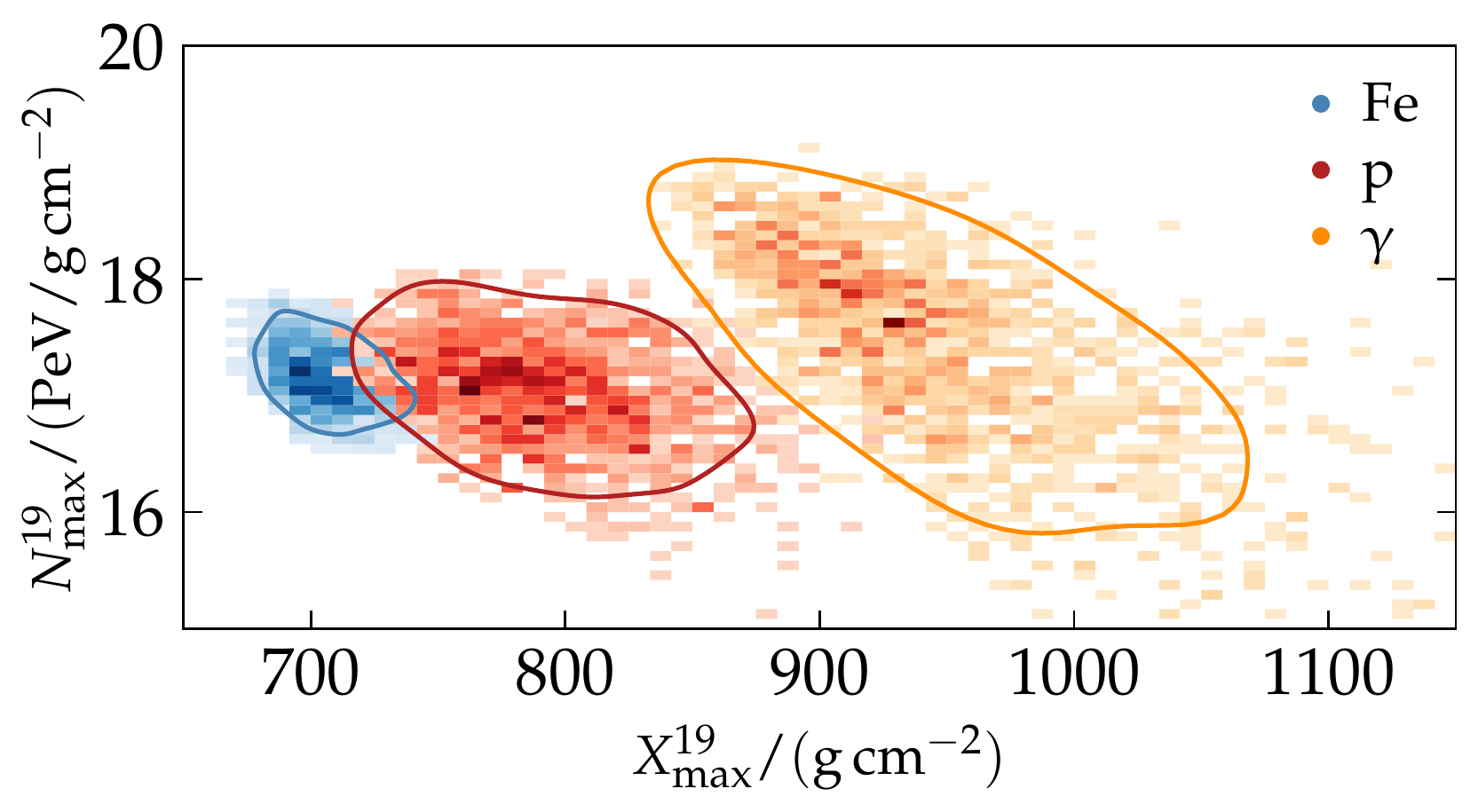}
    \caption{Two-dimensional distribution of the best-fit values of $\Xmax^{19}$ and $N_\text{max}^{19}$ using the GH function to fit simulated longitudinal profiles of showers with primary energies of $10^{18.5}\,\eV$, $10^{19}\,\eV$, and $10^{19.5}\,\eV$.
    All showers were simulated using the \textsc{Sibyll2.3d} model of hadronic interactions.
    The curved lines show the estimated $1\sigma$ extent of the respective distributions.
    }
    \label{fig:xmaxnmax}
\end{figure}

\end{widetext}
\end{document}